\documentclass{aa}
\usepackage{times}
\usepackage{graphicx}
\usepackage{natbib}
\usepackage{amsmath}
\def\msun{{\rm\,M_\odot}}
\def\kms{\mbox{km\,s}^{-1}}
\def\magsqas{\mbox{mag}\,/\,\mbox{arcsec}^2}
\def\spose#1{\hbox to 0pt{#1\hss}}
\def\lta{\mathrel{\spose{\lower 3pt\hbox{$\mathchar"218$}}
     \raise 2.0pt\hbox{$\mathchar"13C$}}}
\def\gta{\mathrel{\spose{\lower 3pt\hbox{$\mathchar"218$}}
     \raise 2.0pt\hbox{$\mathchar"13E$}}}
\def\nsim{\sim\!}
\def\nsimeq{\simeq\!}

\begin{document}

\title{The unmixed kinematics and origins of diffuse stellar light in
  the core of the Hydra I cluster (Abell 1060) }

   \subtitle{}

   \author{G. Ventimiglia \inst{1,2}
          \and M. Arnaboldi \inst{2,3}
          \and O. Gerhard \inst{1}
          }

   \offprints{G. Ventimiglia, e-mail: gventimi@mpe.mpg.de}

   \institute{Max-Plank-Institut f\"ur Extraterrestrische Physik,
     Giessenbachstra$\beta$e 1, D-85741 Garching bei M\"unchen, Germany.
     \and European Southern Observatory, Karl-Schwarzschild-Stra$\beta$e 2,
          85748 Garching bei M\"unchen, Germany.
     \and INAF, Osservatorio Astronomico di Pino Torinese, I-10025 Pino Torinese, Italy.
}

   \date{Received October 22, 2010;
     accepted December 16, 2010}

% \abstract{}{}{}{}{} 
% 5 {} token are mandatory
  \abstract
  {Diffuse intracluster light (ICL) and cD galaxy halos are believed
    to originate from galaxy evolution and disruption in clusters.  }
  {The processes involved may be constrained by studying the dynamical
    state of the ICL and the galaxies in the cluster core. Here we
    present a kinematic study of diffuse light in the Hydra I (Abell
    1060) cluster core, using planetary nebulas (PNs) as tracers.  }
  {We used multi-slit imaging spectroscopy with FORS2 on VLT-UT1 to
    detect 56 PNs associated with diffuse light in the central
    $100\times 100\;\mbox{kpc}^2$ of the Hydra I cluster, at a
    distance of $\nsim50$ Mpc. We measured their [OIII] $m_{5007}$
    magnitudes, sky positions, and line-of-sight velocity distribution
    (LOSVD), and compared with the phase-space distribution of nearby
    galaxies.  }
  {The luminosity function of the detected PNs is consistent with that
    expected at a distance of $\nsim50$ Mpc. Their number density is
    $\nsim4$ times lower for the light seen than expected, and we
    discuss ram pressure stripping of the PNs by the hot intracluster
    medium as one of the possible explanations.  The LOSVD histogram
    of the PNs is highly non-Gaussian and multipeaked: it is
    dominated by a broad central component with $\sigma \nsim
    500\,\kms$ at around the average velocity of the cluster, and
    shows two additional narrower peaks at 1800$\,\kms$ and
    5000$\,\kms$.  The main component is broadly consistent with
    the outward continuation of the intracluster halo of NGC 3311,
    which was earlier shown to have a velocity dispersion of
    $\nsim470\,\kms$ at radii of $\gta50''$. Galaxies with velocities
    in this range are absent in the central
    $100\times100\;\mbox{kpc}^2$ and may have been disrupted earlier
    to build this component. The PNs in the second peak in the LOSVD
    at $5000\,\kms$ are coincident spatially and in velocities with a
    group of dwarf galaxies in the MSIS field. They may trace the
    debris from the ongoing tidal disruption of these galaxies.  }
  {Most of the diffuse light in the core of Abell 1060 is still not
    phase-mixed. The build-up of ICL and the dynamically hot cD halo
    around NGC~3311 are ongoing, through the accretion of material
    from galaxies falling into the cluster core and tidally
    interacting with its potential well.  }

\titlerunning{Kinematics and origins of diffuse light in
  Hydra I cluster core}

%   \titlerunning{Kinematics and origins of diffuse stellar light in
%     the core of the Hydra I cluster}

   \authorrunning{Ventimiglia et al. }

   \maketitle

\section{Introduction}
Intracluster light (ICL) consists of stars that fill up the
cluster space among galaxies and that are not physically bound to any
galaxy cluster members. For clusters in the nearby universe, the morphology
and quantitative photometry of the ICL have been studied with deep
photometric data or by detection of single stars in large areas of sky.

Deep large-field photometry shows that ICL is common in clusters of
galaxies and it has morphological structures with different angular
scales. The fraction of light in the ICL with respect to the total
light in galaxies is between 10\% and 30\%, depending on the cluster
mass and evolutionary status
\citep{Feldmeier04,Adami05,Mihos05,Zibetti05,Krick+Bernstein07,Pierini08}.
The detection of individual stars associated with the ICL, such as
planetary nebulas (PNs)
\citep{Arnaboldi04,Aguerri05,Gerhard07,Castro-Rodriguez+09}, globular
clusters (GCs) \citep{Hilker02,Lee+10}, red giants stars
\citep{Durrell02,Williams07}, and supernovae \citep{Gal-Yam03,Neill05}
is a complementary approach to deep photometry for studying the ICL,
also enabling kinematic measurements for this very low surface
brightness population.

An important open question is the relation between the ICL and the
extended outer halos of brightest cluster galaxies (BCGs), whether
they are independent components or whether the former is a radial
extension of the latter. Using a sample of 683 SDSS clusters,
\cite{Zibetti05} found a surface brightness excess with respect to an
inner $\mbox{R}^{1/4}$ profile used to describe the mean profile of
the BCGs, but it is not known yet whether this cD envelope is simply
the central part of the cluster's diffuse light component or whether
it is distinct from the ICL and part of the host galaxy
\citep{Gonzalez05}.

Both the ICL and the halos of BCGs are believed to have formed from
stars that were tidally dissolved from their former host galaxies or
from entirely disrupted galaxies. A number of processes have been
discussed, starting with early work such as
\citet{Richstone76,Hausman+Ostriker78}. Contributions to the ICL are
thought to come from weakly bound stars generated by interactions in
galaxy groups, subsequently released in the cluster's tidal field
\citep{Rudick06,Rudick09,Kapferer+10}, interactions of galaxies with
each other and with the cluster's tidal field
\citep{Moore+96,Gnedin03,Willman04}, and from tidal dissolution of
stars from massive galaxies prior to mergers with the BCG
\citep{Murante07,Puchwein+10}. Stars in BCG halos may have originated
in both such major mergers as well as through minor mergers with the
BCG. Which of these processes are most important is still an open
issue.

Kinematic studies of the ICL and the cD halos are instrumental in
answering these questions.  The kinematics of the ICL contains the
fossil records of past interactions, due to the long dynamical
timescale, and thus helps in reconstructing the processes that
dominate the evolution of galaxies in clusters and the formation of
the ICL \citep{Rudick06,Gerhard07,Murante07,ArnaGer2010}.  The
kinematics in the cD halos can be used to separate cluster from galaxy
components, as shown in simulations \citep{Dolag10}; so far, however,
the observational results are not unanimous: in both NGC 6166 in Abell
2199 \citep{Kelson02} as well as NGC 3311 in Abell 1060
\citep{Ventimiglia10b} the velocity dispersion profile in the outer
halo rises to nearly cluster values, whereas in the Fornax cD galaxy
NGC 1399 \citep{McNeil+10} and in the central Coma BCGs
\citep{Coccato10a} the velocity dispersion profiles remain flat, and
in M87 in Virgo \citep{Doherty09} it appears to fall steeply to the
outer edge. Evidently, more work is needed both to enlarge the sample
and to link the results to the evolutionary state of the host
clusters.

The aim of this work is to further study the NGC~3311 halo, how it
blends into the ICL, and what is its dynamical status. NGC~3311 is the
cD galaxy in the core of the Hydra I (Abell 1060) cluster.  Based on
X-ray evidence, the Hydra I cluster is the prototype of a relaxed
cluster \citep{Tamura00,Furusho01,Christlein03}.  Surface photometry
is available in the Johnson B, Gunn g and r bands
\citep{vasterberg91}, and the velocity dispersion profile has been
measured out to $\nsim 100"$ \citep{Ventimiglia10b}, showing a steep
rise to $\nsim 470\,\kms$ in the outer halo.  Here we use the
kinematics of PNs from a region of $100 \times 100\;\mbox{kpc}^2$
centered on NGC 3311, to extend the kinematic study to larger radii
and characterize the dynamical state of the outer halo and of the
cluster core.

In Section~\ref{Hydra} we summarize the properties of the Hydra~I
cluster from X-ray and optical observations.  In Section~\ref{MSIS} we
discuss PNs as kinematical and distance probes, and the ``Multi-Slit
Imaging Spectroscopy - MSIS'' technique for their detection in
clusters in the distance range $40-100 {\rm Mpc}$.  We present the
observations, data reduction, identification, and photometry in
Sections~\ref{Observations} and \ref{data}.  In Section~\ref{PNhydra}
we describe the spatial distribution, line-of-sight (LOS) velocity
distribution (LOSVD), and magnitude-velocity plane of the PN sample.
In Section~\ref{Simulation} we use the properties of the planetary
nebulae luminosity function (PNLF) and a kinematic model for the PN
population to predict its LOSVD in MSIS observations. The simulation
allows us to interpret the observed LOSVD and also to determine the
luminosity-specific PN number or
$\alpha$ parameter for the halo of NGC 3311. In Section~\ref{clusub}
we correlate the velocity subcomponents in the PN LOSVD with kinematic
substructures in the Hydra~I galaxy distribution and discuss
implications for galaxy evolution and disruption in the cluster
core. Finally, Section~\ref{Conclu} contains a summary and the
conclusions of this work.

\section{The Hydra~I cluster of galaxies (Abell 1060)} \label{Hydra}
The Hydra~I cluster (Abell 1060) is an X-ray bright, non-cooling flow,
medium compact cluster in the southern hemisphere, whose central
region is dominated by a pair of non-interacting giant elliptical
galaxies, NGC~3311 and NGC~3309. NGC~3309 is a regular giant
elliptical (E3) and NGC~3311 is a cD galaxy with an extended halo
\citep{vasterberg91}.

{\it X-ray properties of Hydra~I} - Except for two peaks associated
with the bright elliptical galaxies NGC~3311 and NGC~3309, the X-ray
emission from the hot intracluster medium (ICM) in the Hydra I
(A~1060) cluster is smooth and lacks prominent spatial
substructures. The center of the nearly circularly symmetric emission
contours roughly coincides with the center of NGC~3311
\citep{Tamura00,Yamasaki02,Hayakawa04,Hayakawa06}.  A faint extended
emission with angular scale $<1'$ trailing NGC 3311 to the northeastern,
overlapping with an Fe excess, could be due to gas stripped from NGC
3311 if the galaxy moved towards the south-west with velocity $\gta
500\,\kms$, according to \citet{Hayakawa04, Hayakawa06}.  The total gas
mass and iron mass contained in this region are $\nsim 10^9\msun$ and
$2\times 10^7\msun$, respectively \citep{Hayakawa04,Hayakawa06}. The
emission components of NGC~3311 and NGC~3309 themselves are small,
extending to only $\nsim10"\nsimeq 2.5$ kpc, suggesting that both
galaxies lost most of their gas in earlier interactions with the
ICM. In both galaxies, the X-ray gas is hotter than the equivalent
temperature corresponding to the central stellar velocity dispersions,
and in approximate pressure equilibrium with the ICM
\citep{Yamasaki02}.

On cluster scales the X-ray observations show that the hot ICM has a
fairly uniform temperature distribution, ranging from about
$3.4\,\mbox{KeV}$ in the center to $2.2\,\mbox{KeV}$ in the outer
region, and constant metal abundances out to a radius of
$230$ kpc. Deviations from uniformity of the hot gas temperature and
metallicity distribution in Hydra~I are in the high metallicity region
at $\nsim1.5\,\mbox{arcmin}$ northeastern of NGC~3311, and a region at a
slightly higher temperature at 7 arcmin south-east of NGC~3311
\citep{Tamura00,Furusho01,Yamasaki02,Hayakawa04,Hayakawa06,Sato07}. Based
on the overall regular X-ray emission and temperature profile, the
Hydra~I cluster is considered as the prototype of an evolved and
dynamically relaxed cluster, with the time elapsed since the last
major subcluster merger being at least several Gyr. From the X-ray
data the central distribution of dark matter in the cluster has been
estimated, giving a central density slope of $\nsimeq -1.5$ and a mass
within $100$~kpc of $\nsimeq 10^{13}\msun$ \citep{Tamura00,Hayakawa04}.
Given these properties, the Hydra I cluster is an interesting target
for studying the connection between the ICL and the extended halo of
NGC~3311.

{\it The cluster average velocity and velocity dispersion} - From a
deep spectroscopic sample of cluster galaxies extending to
$\mbox{M}_R\leq-14 $, \cite{Christlein03} derive the average cluster
redshift (mean velocity) and velocity dispersion of Hydra I. We adopt
their values here: $\bar{v}_{\rm Hy}=3683\pm46\,\kms$, and
$\sigma_{\rm Hy}=724\pm31\,\kms$. The sample of measured galaxy
spectra in Hydra I is extended to fainter magnitudes $\mbox{M}_V >
-17$ through the catalog of early-type dwarf galaxies published by
\cite{Misgeld08}; their values for the average cluster velocity and
velocity dispersion are $\bar{v}_{\rm Hy}=3982\pm 148\,\kms$ and
$\sigma_{\rm Hy}=784\,\kms$, with the average cluster velocity at
somewhat higher value with respect to the measurement by
\cite{Christlein03}.  Both catalogs cover a radial range of $\nsim
300\, \mbox{kpc}$ around NGC 3311. Close to NGC 3311, a predominance of
velocities redshifted with respect to $\bar{v}_{\rm Hy}$ is seen, but
in the radial range $\nsim 50-300\, \mbox{kpc}$, the velocity
distribution appears well-mixed with about constant velocity
dispersion.

{\it Distance estimates} - The distance to the Hydra~I cluster is not
well constrained yet, as different techniques provide rather different
estimates. The cosmological distance to Abell 1060 based on the
cluster redshift is $51.2\pm5.7\,\mbox{Mpc}$ assuming $H_0 = 72 \pm 8
$km$^{-1}$ Mpc$^{-1}$ \citep{Christlein03}, while direct measurements
using the surface brightness fluctuation (SBF) method for 16
galaxies give a distance of $ 41\,\mbox{Mpc}$ \citep*{Mieske05}.

The relative distance of NGC~3311 and NGC~3309 along the line of sight
is also controversial. Distance measurements based on the globular
cluster luminosity function locate NGC~3311 about $10 \,\mbox{Mpc}$ in
front of NGC~3309, which puts NGC 3309 at $61 \,\mbox{Mpc}$
\citep{Hilker03}, while SBF measurements suggest the opposite, with
NGC~3311 now at shorter distance of about $41 \,\mbox{Mpc}$ and
NGC~3309 even closer at $36 \,\mbox{Mpc}$, $5 \,\mbox{Mpc}$ in
front of NGC~3311 \citep{Mieske05}.

In this work we assume a distance for NGC 3311 and the Hydra~I cluster
of $51\,\mbox{Mpc}$, corresponding to a distance modulus of $33.54$.
Then $1"$ corresponds to $0.247\,\mbox{kpc}$. The systemic velocity
for NGC~3311 and its central velocity dispersion are $\mbox{v}_{\rm
  N3311}=3825$ ($3800$) $\pm8\,\kms$ (heliocentric; without and in
brackets with relativistic correction), and $\sigma_0= 154\pm 16
\,\kms$ \citep{Ventimiglia10b}. The systemic velocity of NGC~3309 is
$\mbox{v}_{\rm N3309}=4099\,\kms$ \citep{Misgeld08}. The velocities of
the other Hydra~I galaxies are extracted from the catalogs of
\cite{Misgeld08} and \cite{Christlein03}.

\section{Probing the ICL kinematics using planetary nebulas}\label{MSIS}

\subsection{Planetary nebulas as kinematical probes and distance 
indicators}

PNs occur as a brief phase during the late
evolution of solar-type stars. In stellar populations older than 2
Gyrs, about one star every few million is expected to be in the PN
phase at any one time \citep{Buzzoni06}. Stars in the PN phase can be
detected via their bright emission in the optical [OIII]$\lambda$ 5007
\AA\ emission line, because the nebular shell re-emits $\nsim10\%$ of
the UV photons emitted by the stellar core in this single line
\citep{Ciardullo05}. When the [OIII]  emission line is detected, the
line-of-sight velocity of the PN can be easily measured.

The number density of PNs traces the luminosity density of the parent
stellar population. According to single stellar population theory, the
luminosity-specific stellar death rate is independent of the precise
star formation history of the associated stellar population
\citep{RenBuzzoni86, Buzzoni06}. This property is captured in a simple
relation such that
\begin{equation}\label{totalPN}
N_{PN} = \alpha L_{gal}
\end{equation}
where $N_{PN}$ is the number of all PNs in a stellar population,
$L_{gal}$ is the bolometric luminosity of that parent stellar
population and $\alpha$ is the luminosity-specific PN number.  The
predictions from stellar evolution theory are further supported by
empirical evidence that the PN number density profiles follow light in
late- and early-type galaxies \citep{Kimberly08, Coccato09}, and that
the luminosity-specific PN number $\alpha$ stays more or less constant
with (B-V) color. The empirical result that the rms scatter of
$\alpha$ for a given color is about a factor 2-3 remains to be
explained, however \citep{Buzzoni06}.

The planetary nebula luminosity function (PNLF) technique is one of
the simplest methods for determining extragalactic distances. This is
based on the observed shape of the PNLF.  At faint magnitudes, the
PNLF has the power-law form predicted from models of uniformly
expanding shells surrounding slowly evolving central stars
\citep{HenWester63,Jacoby80}. However, observations and simulations
have demonstrated that the bright end of the PNLF dramatically breaks
from this relation and falls to zero very quickly, within $\nsim 0.7$
mag \citep{Ciardullo98,Mendez+Soffner97}. It is the constancy of the
cutoff magnitude, $M^* = - 4.51$, and the high monochromatic
luminosity of PNs, that makes the PNLF such a useful standard candle.

\subsection{The Multi-Slit Imaging Spectroscopy technique}
At the distance of the Hydra~I cluster, the brightest PNs at the PNLF
cutoff have an apparent $m_{5007}$ magnitude equal to $29.0$,
corresponding to a flux in the [OIII]$\lambda5007$\AA\ line of
$\nsim8\times10^{-18}\,\mbox{erg}\,\mbox{s}^{-1}\mbox{cm}^{-2}$
according to the definition of $m_{5007}$ by \cite{Jacoby89}. To
detect these faint emissions we need a technique that substantially
reduces the noise from the night sky.  This is possible by using a
dedicated spectroscopic technique named ``Multi-Slit Imaging
Spectroscopy'' \citep[MSIS,][]{Gerhard05,Arnaboldi07}.

MSIS is a blind search technique that combines the use of a mask of
parallel slits, a dispersing element, and a narrow band filter
centered at the redshifted [OIII]$\lambda5007\mbox{\AA}$ emission
line. With MSIS exposures, PNs and other emission objects in the
filter's wavelength range which happen to lie behind the slits are
detected, and their velocities, positions, and magnitudes can be
measured at the same time. The [OIII] emission line from a PN is
$\nsim30\, \kms$ wide \citep{Arnaboldi08}, so if dispersed with a
spectral resolution $R \nsim 6000$, it falls on a small number of
pixels, depending on the slit width and seeing.

In this work we use MSIS to locate a sample of PNs in the core of the
Hydra~I cluster and measure their velocities and magnitudes.  Our aim
is to infer the dynamical state of the diffuse light in the
cluster core, as described below in Sections~\ref{Simulation} and
\ref{clusub}.

\section{Observations}\label{Observations}
MSIS data for Hydra~I were acquired during the nights of March 26-28,
2006, with FORS2 on UT1, in visitor mode. The FORS2 field-of-view
(FoV) is $\nsim 6.8 \times 6.8\,\mbox{arcmin}^2$, corresponding to
$\nsim 100\,\times 100\,\mbox{kpc}^2$ at the distance of the
Hydra~I cluster. The effective field area in which it was possible to
position slits with the Grism used here is $44.6\,\mbox{arcmin}^2$.  The
FoV was centered on NGC 3311 at
$\alpha=10\mbox{h}36\mbox{m}42.8\mbox{s}$,
$\delta=-27\mbox{d}31\mbox{m}42\mbox{s}$ (J2000) in the core of the
cluster. The FoV is imaged onto two $2\times2$ rebinned CCDs, with
spatial resolution $0''.252$ per rebinned-pixel. The mask used has
$24\times21$ slits, each $0''.8$ wide and $17''.5$ long. The area
covered with the mask is about $7056\,\mbox{arcsec}^2$, corresponding
to about 4.4 \% of the effective FoV. To cover as much of the field as
possible, the mask was stepped $15$ times so as to fill the distance
between two adjacent slits in the mask. The total surveyed
  area is therefore $29.4\,\mbox{arcmin}^2$, i.e., 66 \% of the
  effective FoV. Three exposures of $800\,\mbox{sec}$ were taken at
each mask position to facilitate the removal of cosmic rays during the
data reduction process.

The dispersing element was GRISM-600B with a spectral resolution of
$0.75\,\mbox{\AA}\, \mbox{pixel}^{-1}$ (or $
1.5\,\mbox{\AA}\,\mbox{rebinned-pixel}^{-1}$) at $5075\,
\mbox{\AA}$. With the adopted slit width, the measured spectral
resolution is $4.5\,\mbox{\AA}$ or $270\,\kms$. Two narrow band
filters were used, centered at $5045\,\mbox{\AA}$ and
$5095\,\mbox{\AA}$, respectively, both with $60\,\mbox{\AA}$
FWHM. This ensures the full coverage of the Hydra~I cluster LOS
velocity range.  Each illuminated slit in the mask produces a
two-dimensional spectrum of 40 rebinned pixels in the spectral
direction and 70 rebinned pixels in the spatial direction.

The seeing during the observing nights was in the range from $0''.6$
to $1''.5$.  For the average seeing ($0''.9$) and with the spectral
resolution of the set-up, monochromatic point-like sources appear in
the final spectra as sources with a total width of $\nsim 5$ pixels in
both the spatial and wavelength directions.

Biases and through-mask flat field images were also taken.  Arc-lamp
calibration frames with mask, Grism and narrow band filter were
acquired for the extraction of the 2D spectra, their wavelength
calibration and distortion correction. Long slit data for the standard
star LTT~7379 with narrow band filter and Grism were acquired for flux
calibration.

\section{Data reduction and analysis} \label{data} The data reduction
is carried out in $IRAF$ as described in \cite{Arnaboldi07} and
\cite{Ventimiglia08}. The frames are registered and co-added after
bias subtraction. The continuum light from the bright galaxy halos is
subtracted using a median filtering technique implemented in the
$IRAF$ task $.images.imfilter.median$, with a rectangular window of
$19 \times 35$ pixels. Then emission line objects are identified, and
2D-spectra around the emission line positions are extracted,
rectified, wavelength and flux calibrated, and background
subtracted. Finally the wavelength of the redshifted [OIII]$\lambda$
5007\AA\ emission line for all the identified sources is measured via
a Gaussian fit. The heliocentric correction for the PN velocities
is $-5.44\,\kms$.

\subsection{Identification of Emission-Line Objects}\label{ident}

All emission line objects found are classified according to the
following criteria as
\begin{itemize} 
\item PN candidates: unresolved emission line objects, both
in wavelength and spatial direction, with no continuum; 
\item background galaxy candidates: unresolved emission line objects
  with continuum or resolved emission line objects both with and
  without continuum.
\end{itemize}
The total number of detected emission line sources in our data set is
82, of which 56 are classified as PN candidates and 26 as background
galaxy candidates, of which 6 are classified as [OII] emitters and the
remaining 20 as candidate Ly$\alpha$ galaxies\footnote{Note
    that the equivalent widths (EWs) of the PN candidates are mostly
    distributed between 30 \AA\ $< EW < 100$ \AA\ , similar to the EWs
    of the background galaxy candidates, and cannot therefore be used
    to discriminate between both types of emission sources.  This is
    because these distant PNs are faint and the continuum level in the
    MSIS images is given by the $1\sigma$ limit from the sky noise;
    see Section~\ref{subsecphot}.}.

For details on the background galaxy candidates see
\cite{Ventimiglia10a}. Note that the background galaxy classification
is independent of luminosity and that these objects have a broad
equivalent width distribution.  Therefore, the fact that the PN
candidates (unresolved emission line objects without detectable
continuum) have a luminosity function as expected for PNs observed
with MSIS at a distance of $\nsim50$ Mpc (see Section~\ref{predict}),
implies that the large majority of these PN candidates must indeed be
PNs.  In addition, Fig.~1 of \citet{Ventimiglia10a} shows that all of
the background galaxy candidates but two fall in the blue filter in
the velocity range between $1000\,\kms$ and $2800\,\kms$, blue-shifted
with respect to the Hydra I cluster.  Several of the {\sl unresolved}
background galaxies in this blue-shifted velocity range have a
continuum level just above the detectability threshold, suggesting
that the PN candidate sample may contain a few background galaxy
contaminants in this velocity range whose continuum is too faint to
detect.

The two background galaxies seen in the red filter are both extended
and have medium bright emission fluxes; one has a very bright
continuum, the other no detectable continuum. From this we conclude
that the residual contamination of the PN candidate sample at
velocities $>3000\,\kms$ must be minimal. With this in mind, we will
in the following simply refer to the PN candidates as PNs.

\subsection{Photometry}
\label{subsecphot}
Magnitudes of the PN candidates are computed using the $m_{5007}$
definition by \cite{Jacoby89}, $m_{5007}=-2.5\log F_{5007}-13.74$,
where $F_{5007}$ is the integrated flux in the line computed in
circular apertures of radius $0''.65-0''.85$ in the 2D spectra,
measured using the $IRAF$ task $.noao.digiphot.aphot.phot$. The
$1\sigma$ limit on the continuum in these spectra is $7.2\times
10^{-20}\mbox{erg}\,\mbox{cm}^{-2}\mbox{s}^{-1}\mbox{\AA}^{-1}$.

\subsubsection{Photometric errors and completeness function}
The photometric errors are estimated using simulations on a sample of
2D wavelength, flux calibrated and background subtracted spectra. For
each simulation 100 artificial PN sources are generated using the
$IRAF$ task $.noao.artdata.mkobject$. The adopted PSF is a Gaussian
with a dispersion obtained by fitting a 2D Gaussian to the profile of
a detected PN candidate with adequate signal-to-noise. The $\sigma$
value is $1.1$ pixels, i.e., $\mbox{FWHM}\nsim0''.7$, and the FWHM in
wavelength is $\nsim 4\, \mbox{\AA}$. The simulated PN samples have
luminosity functions (LFs) given by a delta function at one of five
different input magnitudes (29.3, 29.7, 30.1, 30.5 and 30.9 mag). The
output magnitudes on the 2D spectra are measured with
$.noao.digiphot.aphot.phot$ using circular apertures, in the same way
as for real PN candidates. In these experiments, no significant
systematic shift in the magnitudes was found, and the standard
deviation of the retrieved magnitude distribution is adopted as the
measured error at the respective {\sl output} magnitude.

On the basis of these simulations, we thus model the
errors for the MSIS $m_{5007}$ photometry, which increase
approximately linearly towards fainter magnitudes, by
\begin{equation}\label{magerr}
\epsilon\nsimeq0.25(\mbox{m}_{5007}-28.5)\;\;  [29.0,30.4].
\end{equation}
We then evaluate a completeness correction function, using the
fraction of objects retrieved at each magnitude as these become
fainter. This fraction is nearly 100\% at $29.0\;\mbox{mag}$, the
apparent magnitude of the PNLF bright cutoff at 51 Mpc, and decreases
linearly down to 10-20\% at $30.4\;\mbox{mag}$, the detection limit
magnitude of our observations. We model this dependence by
\begin{equation}
\label{corrfunc}
f_c\simeq
\begin{cases}
1 &\text{if $\mbox{m}_{5007}\leq 29.0$},\\
0.64(-\mbox{m}_{5007}+30.55)    &\text{if $29.0<\mbox{m}_{5007}\leq 30.4$},\\
0 &\text{if $\mbox{m}_{5007}> 30.4$}.
\end{cases}
\end{equation}
The error distribution and the completeness function are used in
Section~\ref{Simulation} below to perform simulations of the LOSVD
for the PN sample.

\section{The PN sample in Hydra~I}\label{PNhydra}

\begin{figure*}[hbt!] \centering
\includegraphics[width=6.0cm]{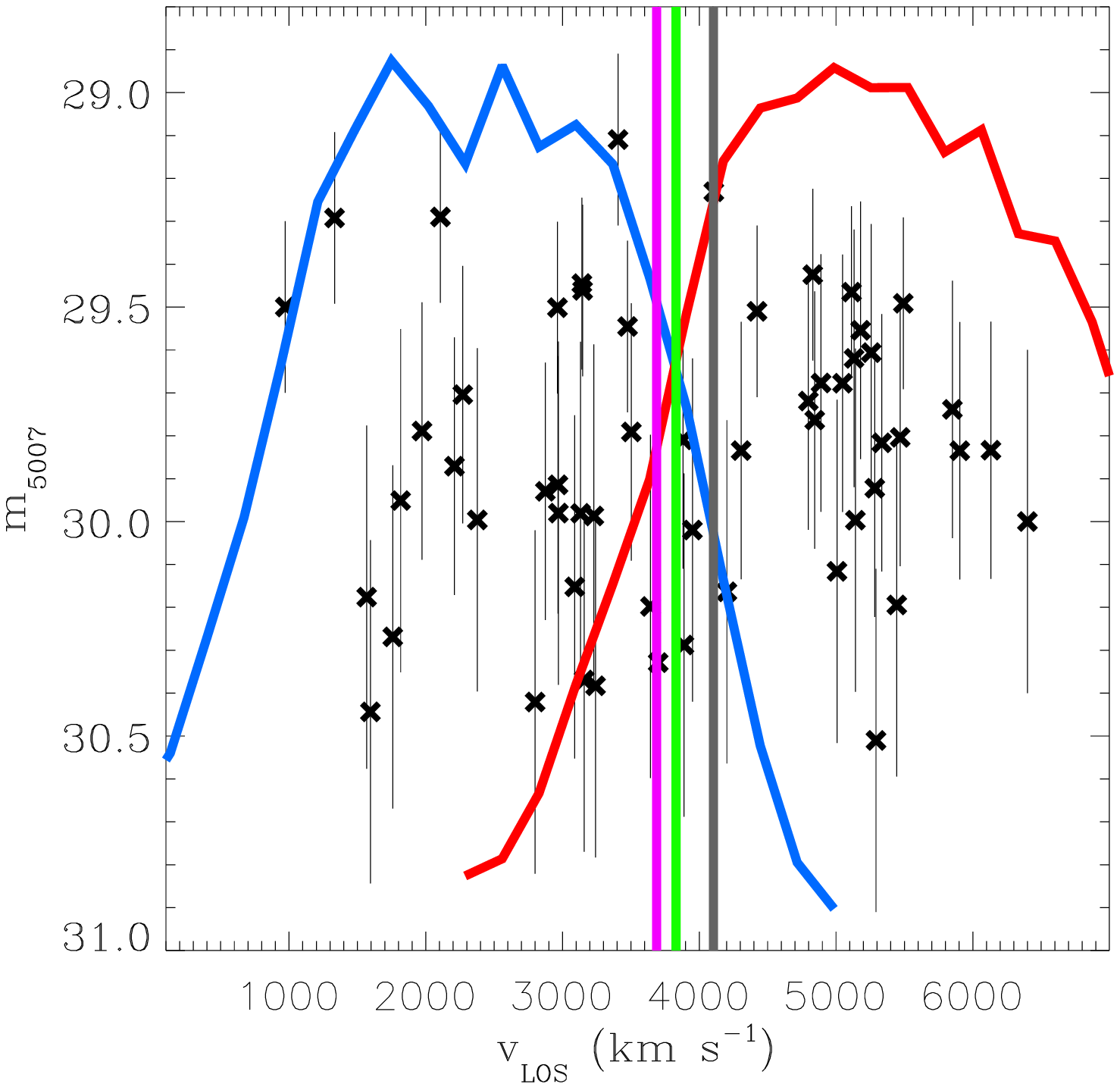}
\includegraphics[width=6.0cm]{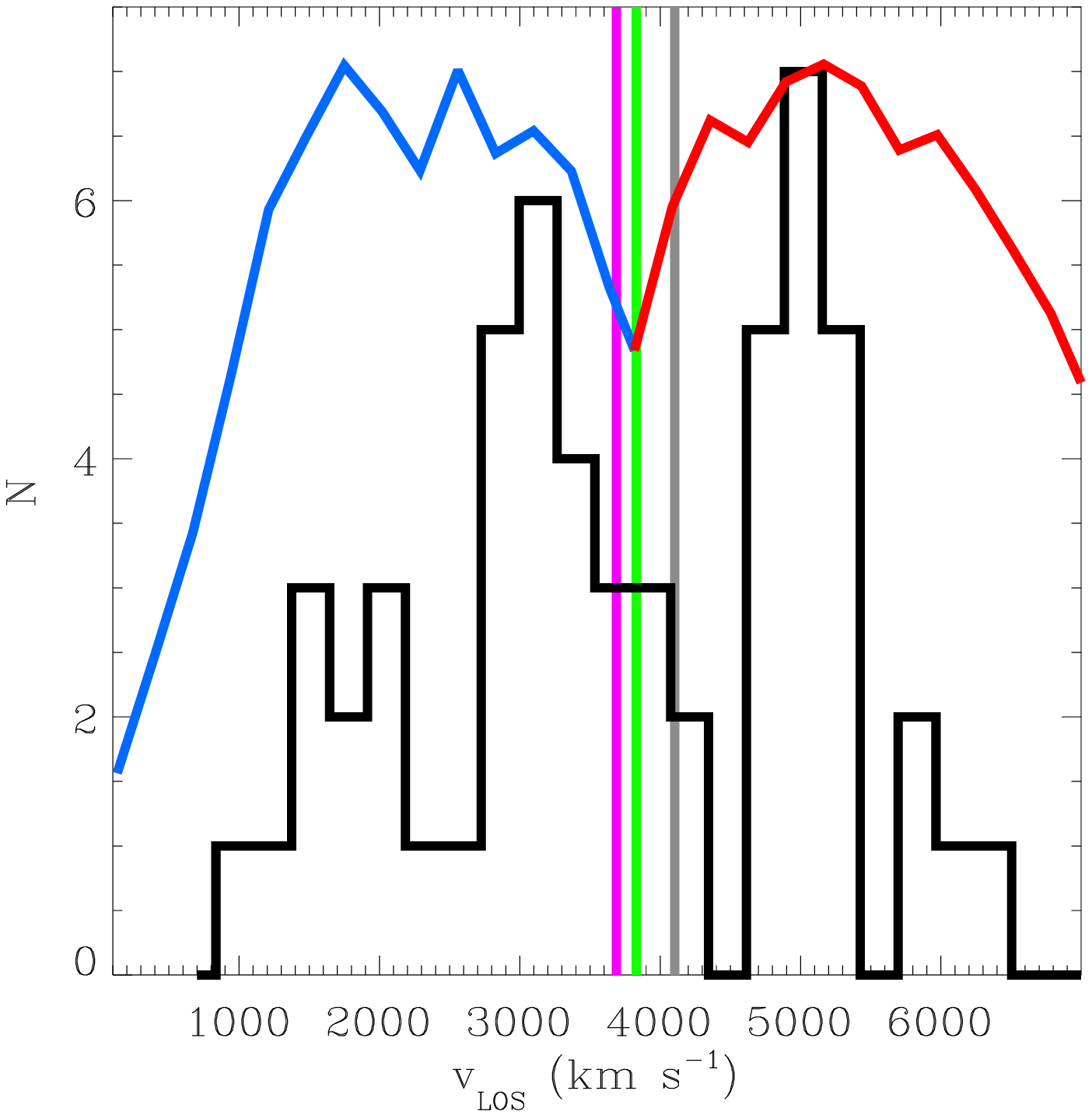}
\includegraphics[width=6.0cm]{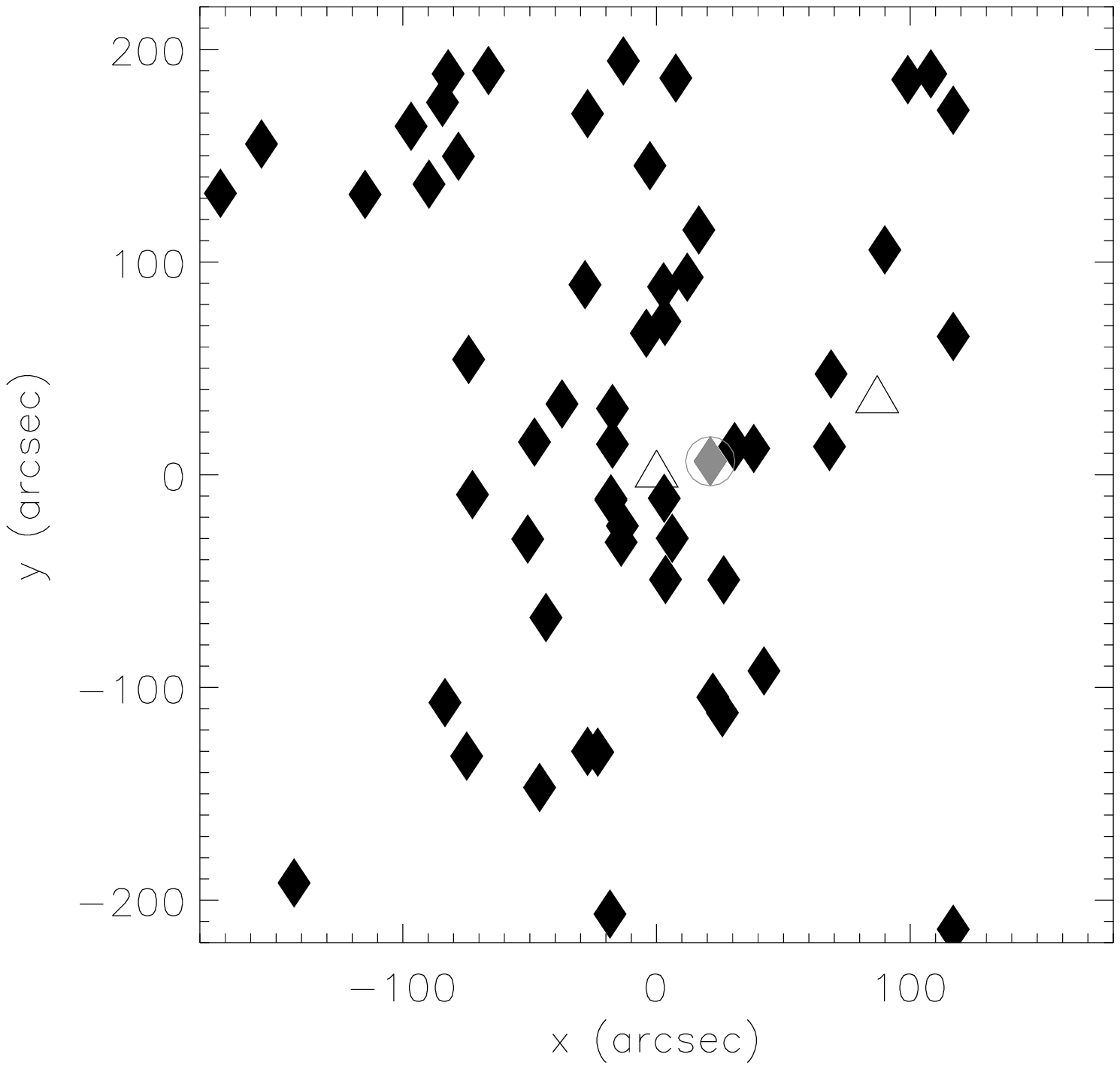}
\caption{PNs in the Hydra I cluster core. \textit{Left panel}: the PN
  velocity-magnitude distribution. The black crosses show the entire
  sample of 56 PN candidates.  The blue and red lines are the measured
  transmission curves of the blue and the red filter, respectively,
  normalized so that the maximum transmission is near the theoretical
  bright cutoff of the PNLF at the distance of Hydra~I.
% (horizontal black line).  
  \textit{Central panel}: the PN LOSVD (black
  histogram). The bins in velocity are $270\,\kms$
  wide. The blue and the red solid lines show again the suitably
  normalized transmission curves of the blue and red filters.  The
  vertical magenta, green and gray lines in both panels mark the
  systemic velocity of Hydra~I, NGC~3311 and NGC~3309, respectively.
  \textit{Right panel}: Spatial distribution of the PNs (black
  diamonds) in the MSIS field. The field is centered on NGC~3311 and
  has size $\nsim 100 \times100\,\mbox{kpc}^2$; north is up and east to
  the left. The two open triangles indicate the positions of NGC~3311
  (center) and NGC~3309 (upper right). The PN indicated by the gray
  symbol is the only object compatible with a PN bound to NGC~3309,
  based on its position on the sky and LOS velocity, $v_{\mbox{ gray
    PN}}=4422\,\kms$.}
\label{velocity-magnitude}
\end{figure*}

Our PN catalog for the central $(100\,\mbox{kpc})^2$ of the Hydra~I
cluster contains 56 candidates, for which we measure $\mbox{v}_{LOS}$,
$x_{PN}$, $y_{PN}$ and $m_{5007}$. The detected PN velocities cover a
range from $970\,\kms$ to $6400\,\kms$ with fluxes from
$2.2\times10^{-18} \mbox{erg}\,\mbox{cm}^{-2}\mbox{s}^{-1}$ to
$7.6\times10^{-18} \mbox{erg}\,\mbox{cm}^{-2}\mbox{s}^{-1}$.  The
detected sample of objects have a magnitude distribution compatible
with the PNLF at the distance of Hydra~I; see also
Section~\ref{predict}.

{\it The magnitude-velocity plane} - The properties of the PN sample
in the velocity-magnitude plane are shown in the left panel of
Fig.~\ref{velocity-magnitude}\footnote{This plot is based on more
  accurate photometry than and updates Fig.1 of
  \cite{Ventimiglia08}.}. In this plot, the apparent magnitude of the
PNLF bright cutoff at the distance of the Hydra~I cluster corresponds
to a horizontal line at $29.0\,\mbox{mag}$. The blue and red lines are
the filter transmission curves, as measured from the spectra,
normalized so that the maximum transmission occurs near the PNLF
bright cutoff.  The PNs are indeed all fainter than $m_{5007}=29.0$
and extend to the detection limit magnitude, $\mbox{m}_{\mbox{ dl}}$. This
is slightly different for the two filters; the faintest PNs detected
through the blue filter have $\mbox{m}_{\mbox{ B,dl}}=30.45$, and those
detected with the red filter have $\mbox{m}_{\mbox{ R,dl}}=30.3$.

{\it The PN LOSVD} - The measured LOSVD of the PN sample is shown by
the black histogram in the central panel of
Fig.~\ref{velocity-magnitude}. The velocity window covered by the two
filters is also shown and the systemic velocities of Hydra~I, NGC~3311
and NGC~3309 (see Section~\ref{Hydra}) are indicated by the magenta,
green and gray vertical lines, respectively. These velocities fall in
the middle of the velocity window allowed by the filters, where both
filters overlap. The mean velocity of all PN candidates is
$\bar{\mbox{v}}_{PNs}=3840\,\kms$ and the standard deviation is
$\mbox{rms}_{PNs}=1390\,\kms$. The distribution is highly non Gaussian
and dominated by several individual components. The main peak appears
in the range of velocities from $2400$ to $4400 \,\kms$ and its
maximum is at $\nsim3100\,\kms$, within $1\sigma_{\mbox{ Hy}}$ of the
systemic velocity of the Hydra~I cluster. In the blue filter velocity
range there is a secondary peak at $\nsim1800\,\kms$ that falls
$2-3\sigma_{\mbox{ Hy}}$ from the systemic velocity of Hydra~I. This blue
peak may contain a few background galaxy contaminants, as discussed in
Section~\ref{ident} above. Finally a red peak at $\nsim5000\,\kms$
within $\nsim 2\sigma_{\mbox{Hy}}$ of the cluster mean velocity is
detected in the velocity interval from $4600$ to $5400 \,\kms$, and
there are some PNs with even higher LOS velocities.

{\it The spatial distribution of the PNs} - The locations of the
detected PNs on the sky are shown in the right panel of
Fig.~\ref{velocity-magnitude}. Their spatial distribution can be
characterized as follows:
\begin{itemize}
 \item most PNs follow an elongated north-south distribution
   approximately centered on NGC~3311;
 \item there is no secondary high density concentration around
   NGC~3309.  Only one PN, indicated by the gray symbol in the right
   plot of Fig.~\ref{velocity-magnitude}, has a combination of
   velocity and position that are compatible with a PN bound to the
   halo of NGC~3309;
 \item a possibly separate concentration of PNs is present in the
   northeastern corner of the field.
\end{itemize}

We summarize our main results so far:
\begin{enumerate}
\item The PN candidates detected in the MSIS field have luminosities
  consistent with a population of PNs at the distance of the Hydra I
  cluster.
\item The distribution of PNs in the MSIS field is centered on NGC~3311.
Only one candidate is consistent with being bound to NGC 3309, even though
NGC 3309 is of comparable luminosity to NGC 3311 and, on account of the
X-ray results (see Section~\ref{Hydra}), is likely located in the inner 
parts of the cluster within the dense ICL, at similar distance from us
as NGC 3311.
\item There is no evidence of a single, well-mixed distribution of
  PNs in the central $100 \,\mbox{kpc}$ of the Hydra I cluster,
  contrary to what one would expect from the dynamically relaxed
  appearance of the X-ray emission.  Instead, the observed PNs
  separate into three major velocity components.
\end{enumerate}

\section{Kinematic substructures and $\alpha$ parameter for the
  observed PN sample in Hydra I: comparison with a simulated MSIS
  model}\label{Simulation}

At this point, we would like to reinforce the last point by comparing
the observed velocity distribution with a simple model. The model is
obtained by assuming a phase-mixed PN population placed at the
distance ($51$~Mpc) and mean recession velocity of NGC 3311, and
simulating its line-of-sight velocity distribution by convolving with
the MSIS instrumental set up.  The velocity dispersion of the PN
population is taken to be $464\,\kms$, the highest value
measured from the long-slit data in \citet{Ventimiglia10b}.  In this
way we can test more quantitatively whether the observed multipeaked
LOSVD for PNs in our field is biased by the MSIS observational
set-up or whether it provides evidence of un-mixed components in the
Hydra~I cluster core.

\subsection{Predicting the luminosity function and LOSVD with MSIS for a 
                       model PN population}\label{predict}

We first characterize the model in terms of the intrinsic luminosity
function and LOSVD of the PN population. Then we describe the steps
required to predict the corresponding $m_{5007}$ magnitude vs.\ LOS
velocity diagram and LOSVD that would be measured with the MSIS set
up. In the next subsection we compare the results obtained with the
observed Hydra~I PN sample.

{\it Model for the intrinsic PN population} - The intrinsic PNLF
can be approximated by the analytical function
given by \cite{Ciardullo89}:
\begin{equation} \label{Ciardullo}
N(m)=C\,e^{0.307\;m}\left[1-e^{3(m^*-m)}\right]
\end{equation}
where $m$ is the observed magnitude, $m^*=29.0$ is the apparent
magnitude of the bright cutoff at the adopted distance of NGC 3311,
and $C$ is a multiplicative factor. The integral of $N(m)$ from $m^*$
to $m^* + 8$ gives the total number of PN associated with the
bolometric luminosity of the parent stellar population ($N_{PN}$ in
Eq.~\ref{totalPN}), and the $C$ parameter can be related to the
luminosity-specific PN number $\alpha$ \citep{Buzzoni06}. For our
model we distribute the magnitudes of a PN population according to a
very similar formula fitted by M\'endez to the results of
\citet{Mendez+Soffner97}.

Next we assume that this PN population is dynamically phase-mixed and
that its intrinsic LOSVD is given by a Gaussian centered on the
systemic velocity of NGC~3311, $\bar{\mbox{v}}$,
\begin{equation}
\mbox{G}(\mbox{v})=
\frac{1}{\sigma_{\mbox{ core}}\sqrt{2\pi}}
\exp{\left[\frac{(\mbox{v}-\bar{\mbox{v}})^2}{2\sigma_{\mbox{ core}}^2}\right]}
\label{Gaussian} 
\end{equation}
where here we adopt $\bar{\mbox{v}}=3830\,\kms$ \citep[]
[corrected to the filter frame]{Ventimiglia10b}, and for the velocity
dispersion we take $\sigma_{\mbox{ core}}=464\,\kms$, the highest value
measured from the long-slit data in this paper. This approximates the
velocity dispersion for the intracluster component in the outer halo
of NGC 3311, at central distances $\nsim 20-30 \,\mbox{kpc}$
\citep{Ventimiglia10b}. We will consider the magnitude-velocity
diagram and the LOSVD as histograms in velocity; then in each velocity
bin $\Delta\mbox{v}_i$, the number of PNs is
\begin{equation}
LF(\mbox{v}_i)\simeq N(m) \, G(\mbox{v}_i) \, \Delta\mbox{v}_i
\end{equation}
where $\mbox{G}({\mbox{v}})$ is normalized so that $\sum_i
 \mbox{G}(\mbox{v}_i)\,\Delta\mbox{v}_i=1$.

 {\it Simulating the MSIS observations} - The magnitude-velocity
 diagram for such a model population is modified by a number of
 effects in the MSIS observations, which we simulate as described
 below. The MSIS simulation procedure implements the following steps:
\begin{itemize}
\item the through-slit convolution of the PNLF;
\item the convolution with the filter transmission;
\item the photometric error convolution;
\item the completeness correction;
\item the computation of the LOSVD. 
\end{itemize}

{\it The through-slit PNLF} - The MSIS technique is a blind survey
technique. Therefore the positions of the slits on the sky are not
centered on the detected objects, and the further away an object is
from the center of its slit, the fainter it becomes. This effect is a
function of both seeing and slit width, and it modifies the functional
form of the PNLF, which needs to be accounted for when using the LF
from MSIS PN detected samples. 

In principle, some PNs may be detected in two adjacent slits of the
mask, and this would need to be corrected for. However, at the depth
of the present Hydra I survey this effect is not important for the
predicted PNLF, and indeed no such object has been found in the
sample.

Given a ``true'' PNLF $\mbox{LF}(m)$, the ``{\it through slit PNLF}''
$\mbox{sLF}(m)$ can easily be computed, and depends on slit width and
seeing; for further details see Gerhard et al.\ (2011, in
preparation). The effect of the through-slit correction is to
  shift the $\mbox{sLF}(m)$ faintwards in the observable bright part,
  compared to the ''true'' PNLF.

{\it Convolution with filter transmission} - When the filter
transmission $\mbox{T}(\mbox{v}_i)$ is less then $1$ ($100\%$), it
shifts the through-slit PNLF to fainter magnitudes. The $\Delta m$
depends on the value of the filter transmission curve at the
wavelength $\lambda$ or equivalent binned velocity $\mbox{v}_i$, and
is equal to $\Delta m(\mbox{v}_i)=-2.5\log\mbox{T}(\mbox{v}_i)$.  The
resulting {\sl instrumental PNLF}, the distribution of source
magnitudes before detection, becomes velocity dependent, i.e., 
$\mbox{iLF}(m,\mbox{v}_i)$.

For the present MSIS Hydra~I observations, the combined filter
transmission curve from both filters is defined as
\begin{equation}\label{filttra}
  \mbox{T}(\mbox{v})_i=
   {\mbox{ max}}[\mbox{T}_{\mbox{ B}}(\mbox{v}_i),\mbox{T}_{\mbox{ R}}(\mbox{v}_i)],
\end{equation}
where B and R denote the blue and red filters.  It is $1$ where the
transmission is 100\%, approximately from $\nsim 1500\,\kms$
to $\nsim3300\,\kms$ and from $\nsim 4200\,\kms$
to $\nsim 6300\,\kms$; it is $< 1$ in the filter gap around
$\nsim 3800\,\kms$ and at the low and high velocity ends of
the observed range.

{\it Photometric error convolution} - Once the instrumental LF
$\mbox{iLF}(m,\mbox{v}_i)$ is computed, it must be convolved with the
photometric errors which, for the case of the Hydra I observations,
are given by the linear function in Eq.~\ref{magerr}. Because of the
photometric errors, PNs that are intrinsically fainter than the
detection limit (here $\mbox{mag}\nsim30.4$) may be detected if they
happen to fall on a positive noise peak on the CCD image, and PNs that
are intrinsically brighter than $\mbox{mag}\nsim30.4$ may be lost from
the sample. Generally, because the through-slit PNLF
  $\mbox{sLF}(m)$ increases towards fainter magnitudes, the
  photometric errors scatter more faint objects to brighter magnitudes
  than vice-versa; so the effect of the convolution is to shift the
  PNLF to brighter magnitudes again.

{\it Completeness correction} - The completeness correction at a given
observed magnitude is a multiplicative function which accounts for the
decreasing fraction of PNs at fainter magnitudes detected against the
noise on the MSIS image.  For the case at hand it is given in
Eq.~\ref{corrfunc}. After the last two steps, we arrive at the final
``MSIS PNLF'', $\mbox{MSLF}(m)$ for short.

{\it Computation of the simulated LOSVD} - For each velocity bin the
$\mbox{MSLF}(m,\mbox{v}_i$) is integrated between the apparent
magnitude of the PNLF bright cut off ($m^*=29.0$ for Hydra I) and the
detection limit magnitude in the relevant filter, $m_{\mbox{ f,dl}}$ (see
Section~\ref{PNhydra}), to obtain the ``observed'' cumulative number
of PNs in each velocity bin:
\begin{equation}
N_{\mbox{ MSIS}}(\mbox{v}_i)=\int_{m^*}^{m_{\rm f,dl}}\mbox{MSLF}(m,\mbox{v}_i)dm.
\end{equation}

The most cumbersome step in this procedure is the correction for the
filter transmission, because it makes the final
$\mbox{MSLF}(m,\mbox{v}_i)$ velocity-dependent. It must correctly be
applied {\sl before} the convolution with the photometric errors,
because the latter depend on the flux measured at certain positions on
the CCD. So the errors on the through-slit magnitudes depend on the
filter transmission values of the PNs.

However, we have found that the observed MSLF for the Hydra~I PN
sample, when obtained from wavelength regions where the filter
transmission is $\nsim 100\%$, is very similar to the one obtained by
summing over the entire filter bandpass.  The effect of the velocity
dependence on the overall MSLF must therefore be small, and for the
comparison of simulated and measured LOSVDs below we have therefore
applied the filter transmission only after the error convolution and
completeness correction.

\begin{figure}[h!] \centering
\includegraphics[width=6.5cm]{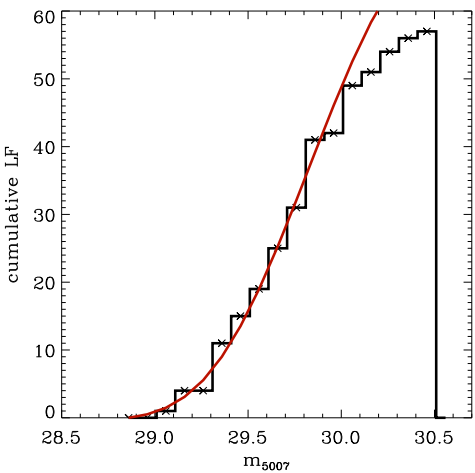}
\caption{Cumulative luminosity function predicted for the present MSIS
  observations and the nominal cutoff magnitude of the Hydra I
  cluster, $29.0$ (full red line, see text), compared with the
  cumulative histogram of the observed $m_{5007}$ magnitudes.}
\label{cmmslf}
\end{figure}

Before we discuss the LOSVD obtained from the complete model, we show
in Fig.~\ref{cmmslf} the predicted cumulative luminosity function
resulting from error convolution, completeness correction, and filter
transmission correction of the through-slit luminosity function,
weighting by the number of observed PNs in each velocity bin. Also
shown in Fig.~\ref{cmmslf} is the cumulative histogram of the
$m_{5007}$ magnitudes for the 56 observed PNs in the MSIS field.  With
a cutoff magnitude of $29.0$ the model fits the observed histogram
fairly well; however, this is not a formal best fit to the
distance. The important point shown by Fig.~\ref{cmmslf} is that the
observed MSIS luminosity function of the PN emission sources in the
Hydra cluster core is evidently consistent with a population of PNs at
$\nsim50$ Mpc distance.

\subsection{Reality of observed kinematic substructures}\label{modvsobs}
\begin{figure}[hbt!] \centering
\includegraphics[width=6.5cm]{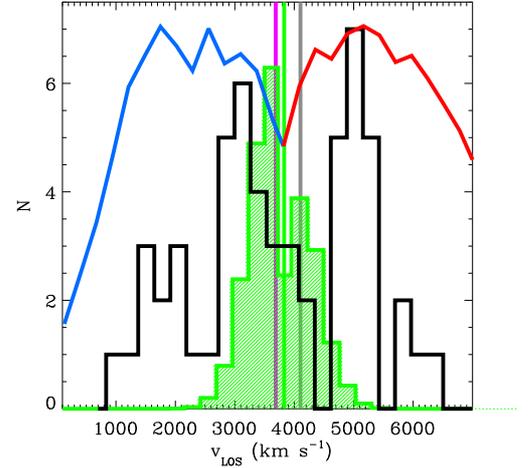}
\caption{LOSVD for the Hydra~I PN sample from
  Fig.~\ref{velocity-magnitude} (black histogram), compared with a
  simulated MSIS LOSVD (green histogram) for a Gaussian velocity
  distribution with $\sigma_{\rm core}=464\,\kms$; see text for further
  details. The blue-red solid line shows the combined filter
  transmission curve as given in Eq.~\ref{filttra}. The vertical
  magenta, green and gray lines mark the systemic velocity of Hydra~I,
  NGC~3311 and NGC~3309, respectively.  }
\label{simulation_histogram}
\end{figure}

The simulated MSIS LOSVD given by $N_{\rm MSIS}(\mbox{v}_i)$ for the
simple Gaussian velocity distribution model and luminosity function of
Eq.~\ref{Ciardullo} is shown as the green histogram in
Fig.~\ref{simulation_histogram}, with the observed PN LOSVD
overplotted in black. The simulated MSIS LOSVD is scaled to
approximately match the observed Hydra~I sample in the central
velocity bins.

The comparison between the simulated LOSVD and the Hydra~I PN LOSVD in
Fig.~\ref{simulation_histogram} identifies the central peak at about
$3100\,\kms$ in the observed PN LOSVD with that of the PN population
associated with the stellar halo around NGC 3311 in the cluster core,
with $\sigma_{\rm core}\nsim 500\,\kms$.  The mean $\bar{\mbox{v}}_{\rm
  core}$ and $\sigma_{\rm core}$ of this component are approximately
consistent with those of the intracluster light halo of NGC~3311
derived from the long-slit kinematic analysis in
\cite{Ventimiglia10b}.  However, the asymmetry and offset of the peak
of the observed histogram (by several $100\,\kms$) relative to the
MSIS convolved model centered at the systemic velocity of NGC 3311
appear significant ($\sigma_{\rm core}/\sqrt{N_{\rm core}}\nsimeq
100\,\kms$), arguing for some real asymmetry of the central velocity
component.  We shall refer to the central peak in the Hydra~I PN
LOSVD in Fig.~\ref{simulation_histogram} as the central ICL component.

Two additional velocity peaks are seen in the LOSVD in
Fig.~\ref{simulation_histogram}, one near $1800\,\kms$ and one at
$\nsim5000\,\kms$, which do not have any correspondence with the
velocity distribution derived for the simulated MSIS model. These
velocity components cannot be explained as artifacts of the MSIS set
up, in particular, the filter gap in the B+R filter combination.  We
will refer to these two velocity components as secondary blue and red
peaks, respectively. They reveal the possible presence of two
kinematical substructures in the core of Abell 1060, whose origins
must be investigated further; see Section~\ref{clusub}.

\subsection{Low $\alpha$-parameter in the core of Hydra~I}\label{reldist}

We now compare the number of observed PNs with the expectations from
the luminosity distribution and kinematics in and around NGC 3311.
One issue is the absence of a clear subcomponent of PNs with velocity
dispersion $\nsim 150-250\,\kms$, as would be expected from the central
$\nsim 25"$ of NGC 3311 \citep{Ventimiglia10b}. It is known that PN
samples in elliptical galaxies are generally not complete in the
central regions because of the increasing surface brightness profile;
PNs are hard to detect against the image noise in the bright centers.
E.g., in observations with the Planetary Nebula Spectrograph, the
threshold surface brightness is typically in the range
$\mu_V=20-22\,\magsqas$ \citep{Coccato09}. In the current Hydra I
data, the PN sample is severely incomplete at $\mu_V=21.0\,\magsqas$
(only two PNs are seen at $\mu_V\nsim 21.0\,\magsqas$, and six at
$\mu_V\gta 21.5\,\magsqas$).  Referring to Fig.~13 of
\citet{Mendez+01}, we estimate that the current sample is not complete
for $\mu_V\lta 22.0\,\magsqas$, which is reached at a distance of
$\nsimeq 30"$ from the center of NGC 3311 (Arnaboldi et al.\ 2011, in
preparation).  At this radius, the projected velocity dispersion has risen
to $\sigma_{N3311}(30")\nsimeq 300-400\,\kms$ \citep{Ventimiglia10b}.
Thus the PNs detected in this paper almost exclusively sample the hot
(intracluster) halo of NGC 3311. The cold inner galaxy component is
not sampled.

The second issue is the observed total number of PNs, given the
detection limit, the instrumental set up and the light in NGC 3311 and
NGC 3309.  Integrating the simulated MSIS luminosity function down to
the detection limit of $30.4\,\mbox{mag}$, we obtain an effective
$\alpha$ parameter for our observations of $\alpha_{\rm MSIS,Hy}=0.012
\alpha_{\rm tot}$, where $\alpha_{\rm tot}$ quantifies the total
number of PNs $8\,\mbox{mag}$ down the PNLF\footnote{This value
  includes the light between adjacent slits for the normalization.}.
This value is similar to $\alpha_{0.5}$, the integrated value
$0.5\,\mbox{mag}$ down the PNLF. It is consistent with
Fig.~\ref{velocity-magnitude}, even though in this figure PNs are seen
up to $1.5\,\mbox{mag}$ fainter than the nominal cutoff magnitude,
because of (i) the shift towards fainter magnitudes due to the slit
losses, and (ii) the completeness correction
(Eq.~\ref{corrfunc}).

\begin{figure}[hbt!] \centering
\includegraphics[width=6.0cm]{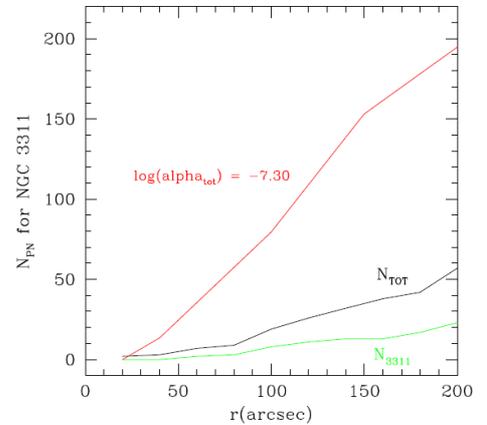}
\caption{ Observed and predicted cumulative PN numbers, as a function
  of radial distance from the center of NGC 3311. The green line shows
  the cumulative number of PNs associated with the central ICL halo of
  NGC 3311, based on their velocities. The black line shows the
  cumulative number of all PNs, without velocity selection. The red
  curve shows the predicted cumulative number of PNs computed using
  the luminosity-specific parameter $\alpha$ estimated as
    explained in the text}, the MSIS observational set-up, and the
  integrated bolometric luminosity in increasing circular apertures
  centered on NGC 3311.
\label{PNcum}
\end{figure}

We can estimate the bolometric $\alpha_{\rm tot}$ for NGC 3311
  from its (FUV-V) color, the relation between (FUV-V) and
  $\log\alpha_{1.0}$ shown in Fig.~12 of \citet{Coccato09}, and
  correcting to $\log\alpha_{\rm tot}$ by using Fig.~8 of
  \citet{Buzzoni06}. The (FUV-V) color is determined from the Galex
  FUV magnitude and the V band magnitude from RC3, both corrected for
  extinction, as described in \citet[][Section 6.1]{Coccato09}.  The
  resulting value, (FUV-V)=6.7, corresponds to $\log\alpha_{1.0}=1.1$
  and $\log\alpha_{\rm tot}=-7.34$.  This is very similar to the value
  of $\log\alpha_{\rm tot}=-7.30$ found for the Fornax cluster cD
  galaxy NGC 1399 \citep{Buzzoni06}.  Using the V band light profile
  of NGC 3311 measured in Arnaboldi et al.\ (2011, in preparation),
  and a bolometric correction of −0.85mag, we can then predict the
  expected cumulative number of PNs within radius R from the center of
  NGC 3311. This is shown as the red curve in Figure~\ref{PNcum},
after subtracting the luminosity within 20'' which is not sampled by
our MSIS observations. Also shown are the cumulative histograms of the
observed number of PNs in the MSIS data, both for all PNs in the
field, and for PNs with velocities in the central velocity component
only.

Fig.~\ref{PNcum} shows that the total number of PNs detected in the
field falls short of the number predicted from the luminosity profile
by a factor $\nsim 4$.  Outside $\nsim100"$, the number of PNs with
velocities consistent with the central ICL halo of NGC 3311 is a
factor $\nsim 2$ lower than the number of all PNs.  Clearly therefore,
some of the light at these radii is in a component different from the
phase-mixed central ICL halo, but the amount is uncertain because we
do not know whether the luminosity-specific $\alpha$-parameter of this
component is similarly low as for the NGC 3311 ICL halo.  For example,
agreement between observed and predicted PN numbers could be achieved
by scaling only the NGC 3311 halo component by a factor $\nsim6$.  On
the other hand, scaling only an outer component will not work, because
the discrepancy in Fig.~\ref{PNcum} is already seen at small
radii. Thus we can conclude that the $\alpha$-parameter of the NGC
3311 ICL halo is low by a factor $4\!-\!6$.

Such an anomalous specific PN number density requires an
  explanation.  One possibility is that the stellar population in the
  halo of NGC 3311 is unusually PN poor; this will need studying the
  stellar population in the galaxy outskirts. A second possibility is
that the ram pressure against the hot X-ray emitting gas in the halo
of NGC 3311 is high enough to severely shorten the lifetime of
the PNs \citep{Dopita+00,Villaver+05}.  In their simulations,
\citet{Villaver+05} consider a gaseous medium of density $n=10^{-4}\,
{\rm cm}^{-3}$ and a relative velocity of $1000 \,\kms$. They find
that the inner PN shell is not significantly affected by the ram
pressure stripping during the PN lifetime, and because the inner
shell dominates the line emission in their model, the PN visibility
lifetime is therefore not shortened relative to an undisturbed PN.
However, with a density of the ICM inside $5'$ around NGC 3311 of
$\nsim 6\times 10^{-3}\, {\rm cm}^{-3}$, and a typical velocity of
$\sqrt{3}\times450\,\kms\nsimeq 800\,\kms$ the ram pressure on the NGC
3311 is $\nsim 40$ times stronger than in their simulated case, so the
ram pressure effects could be much stronger. Unfortunately,
simulations of the evolution of PNs in such dense media are not yet
available, to our knowledge.

If this explanation is correct, PNs should be most efficiently ram
pressure stripped in the innermost, densest regions of the ICM.  Hence
in this case we would expect most of the observed PNs to be located in
the outermost halo of NGC 3311, even those projected onto the inner
parts of our MSIS field. At these outer radii, dynamical time-scales
are longer, and phase-mixing should be less complete. This would fit
well with the unmixed kinematics and spatial distribution of the
observed sample (see also next Section).

The third issue is that we do not see a concentration of PNs around
NGC 3309. As shown in Section~\ref{PNhydra}, only one PN in the
sample, shown by the gray symbol in the right panel of
Fig.~\ref{velocity-magnitude}, has both position and LOS velocity
compatible with being bound to NGC~3309.  Whereas using the relative
total luminosities of NGC 3309 and NGC 3311 to scale the number of PNs
associated with the main LOS velocity component for NGC 3311 in
Fig.~\ref{simulation_histogram} (i.e., 27 PNs), we would expect about
11 PNs associated with the light of NGC 3309 if both galaxies were at
the same distance. There are two possible explanations for this
fact. One is that NGC 3309 is at significantly larger distance than
NGC 3311, such that even PNs at the bright cutoff would be difficult
to see. However, a simple calculation shows that then NGC 3309 would
be put at $\nsim 70$ Mpc well outside the cluster, at variance with
X-ray observations finding that its gas atmosphere is confined by the
ICM pressure (see Section~\ref{Hydra}).  The second possibility is
that, similarly as for NGC 3311, also the PNs in NGC 3309 may be
severely ram pressure stripped by the galaxy's motion through the
dense ICM in the cluster core. This would require that NGC 3309 moves
rapidly through the cluster core, and is physically rather close to
NGC 3311.  Again simulations would be needed to check this
quantitatively.

\section{The substructures in the Hydra I cluster core}\label{clusub}

\begin{figure*}[hbt!] \centering
\includegraphics[width=6cm]{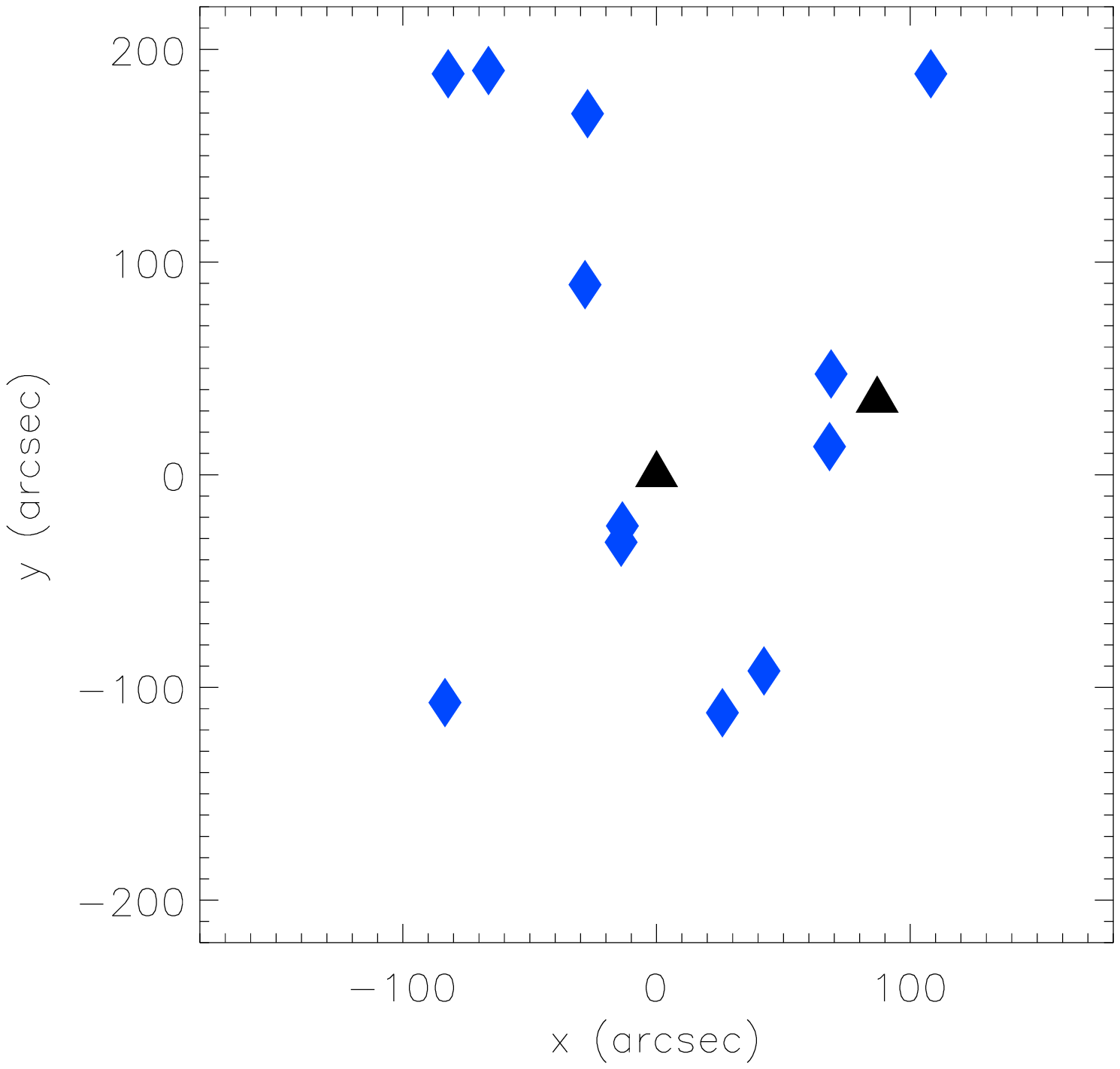}
\includegraphics[width=6cm]{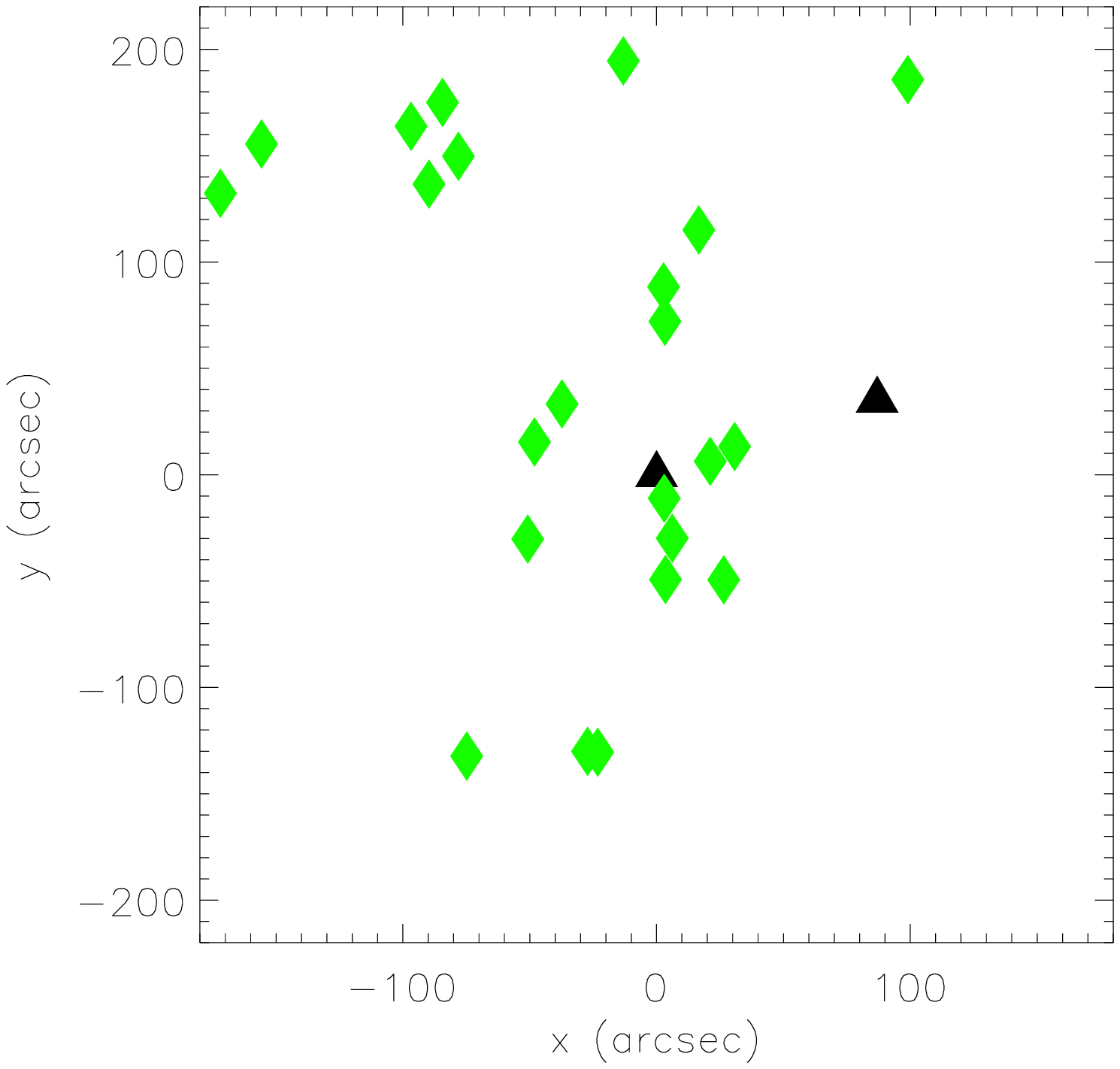}
\includegraphics[width=6cm]{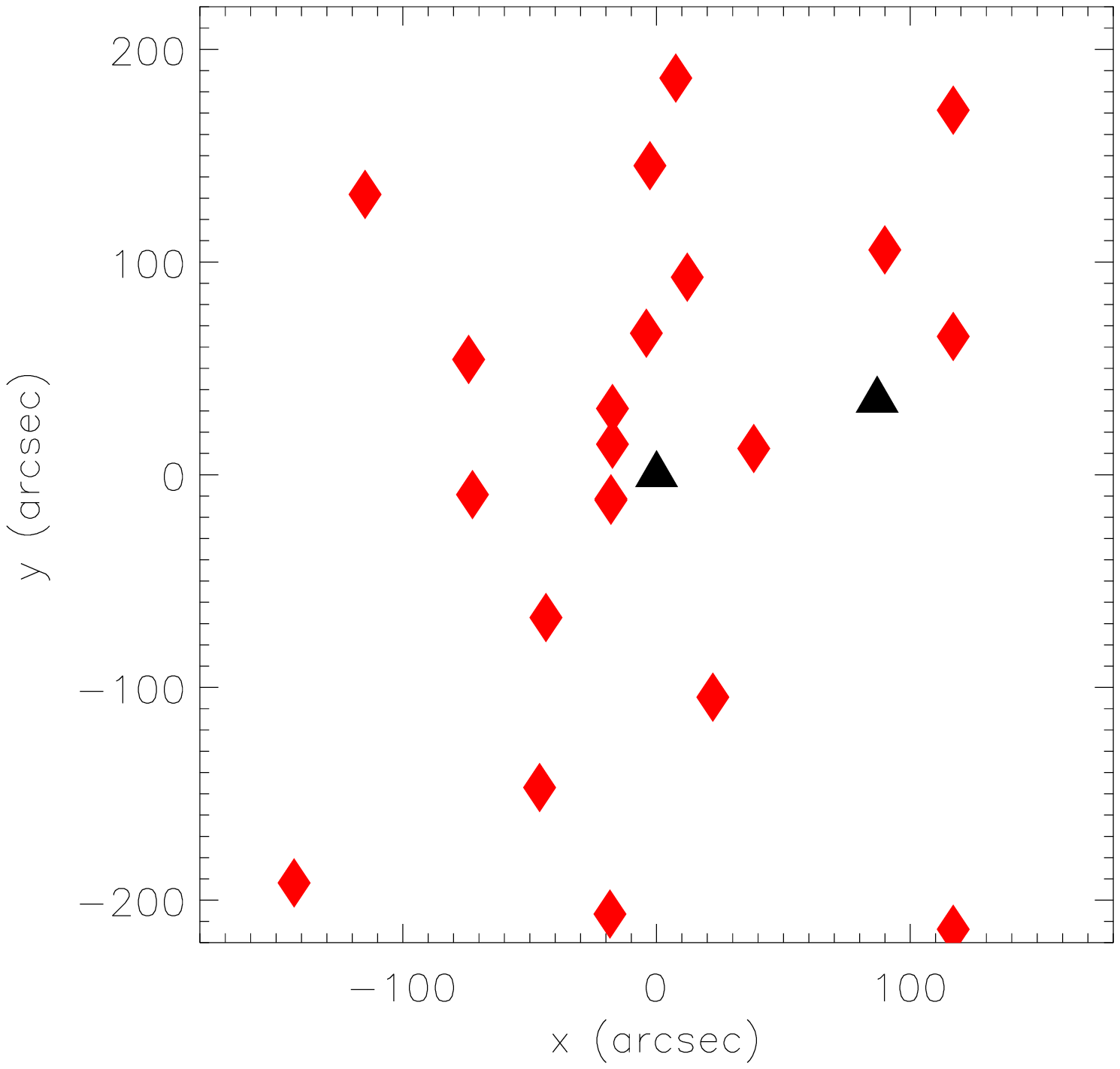}
\caption{\textit{Left panel}: Spatial distribution of the PNs
  associated with the blue secondary peak in the PN LOSVD
  ($<2800\,\kms$).  \textit{Central panel}: Spatial distribution of
  the PNs associated with the central ICL component ( $2800\,\kms$ to
  $4450\,\kms$). \textit{Right panel}: Spatial distribution of the PNs
  associated with the secondary red peak at $ > 4450\,\kms$ in the PN
  LOSVD. The black triangles indicate NGC~3311 (center) and NGC~3309
  (north-west of center), respectively. North is up and east is to the
  left. }
\label{Bphase-space}
\end{figure*}

\begin{figure*}[hbt!] \centering
\setlength{\unitlength}{1cm}
\begin{minipage}[t]{8.4cm}
\mbox{}\\
\begin{picture}(8.0,8.4)
\vspace{0.4cm}
\includegraphics[width=8cm]{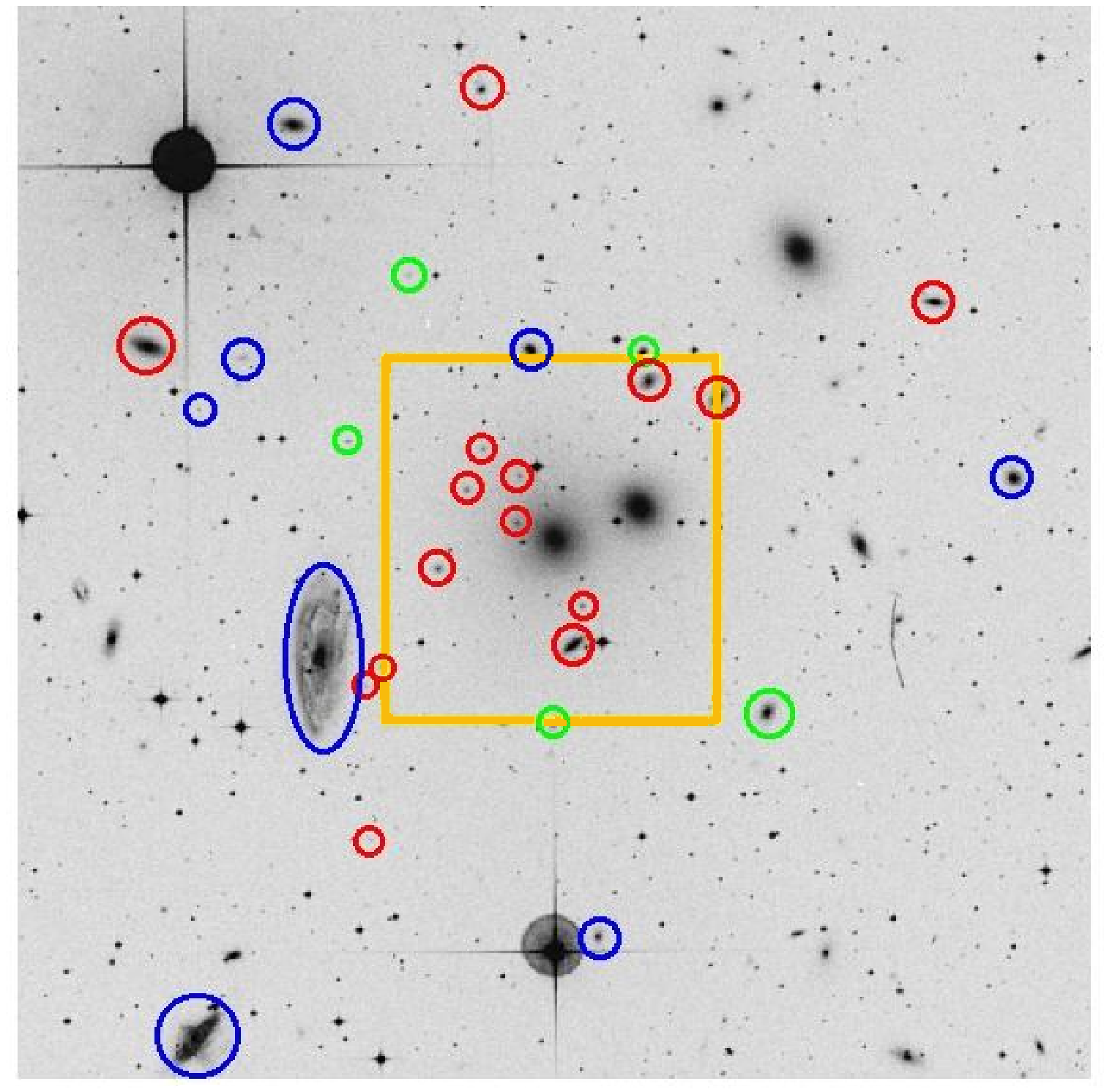}
\end{picture}
\end{minipage}
\hfill
\begin{minipage}[t]{9.5cm}
\mbox{}\\
\begin{picture}(9.5,9.5)
\includegraphics[width=9.5cm]{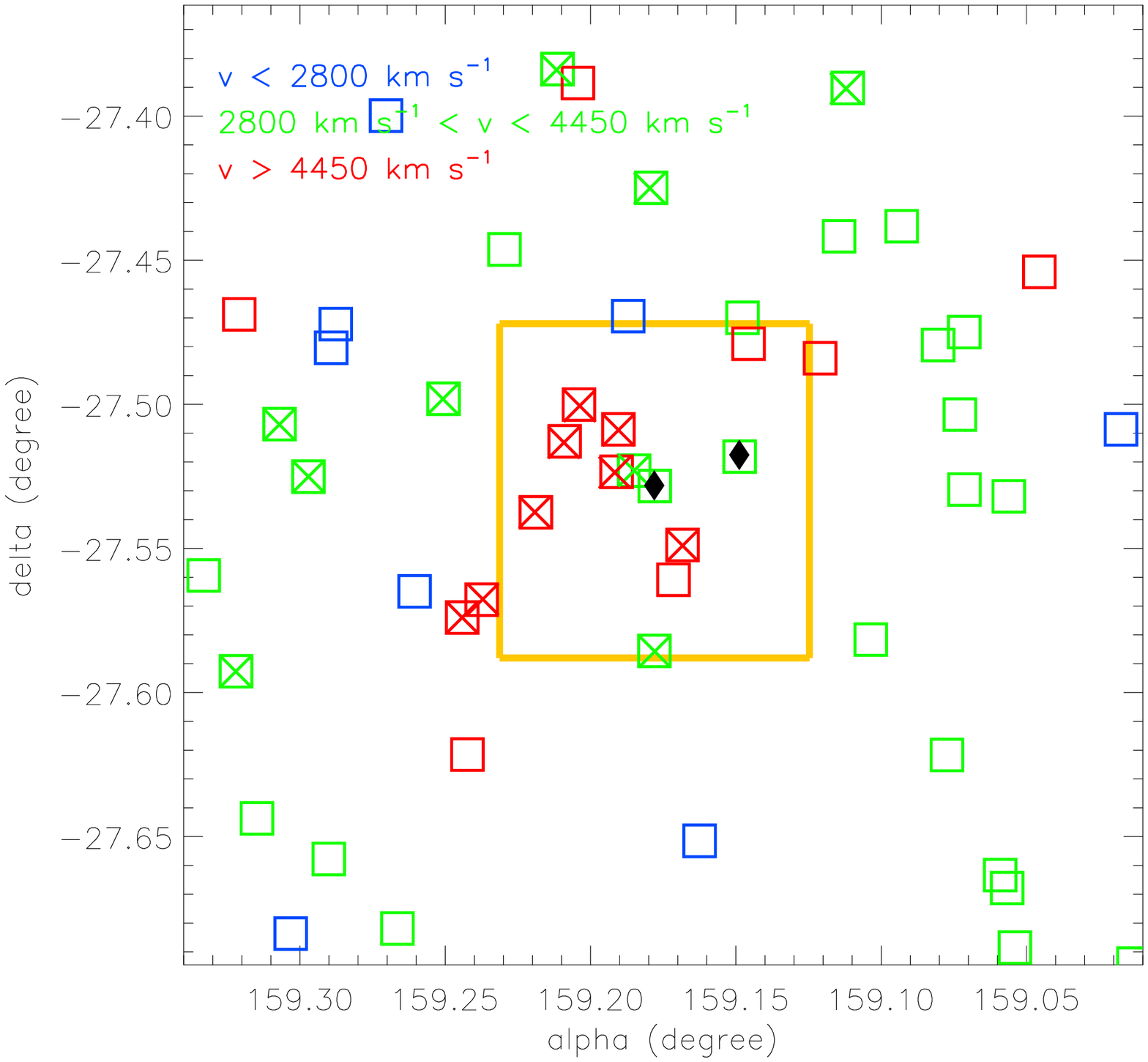}
\end{picture}
\end{minipage}
\caption{\textit{Left panel}: $20 \times 20\,\mbox{arcmin}^2$ DSS
  image of the Hydra~I cluster.  The two bright galaxies at the field
  center are NGC~3311 (center) and NGC~3309 (north-west of center).
  The blue circles indicate galaxies with $v_{sys}<2800\,\kms$, the
  green circles galaxies with $2800\,\kms<v_{sys}<4450\,\kms$ (only
  those within 10 arcmin around NGC~3311 and with $\mbox{m}_R>15.37$),
  and the red circles galaxies with $v_{sys}>4450\,\kms$.
  \textit{Right panel}: Spatial distribution of Hydra~I galaxies in
  the same area of $ 20\,\mbox{arcmin}^2$ centered on
  NGC~3311. Squares indicate galaxies from the catalog of
  \cite{Christlein03} and crosses indicate galaxies from
  \cite{Misgeld08}. The color of the symbols refers to the velocity
  components in the PN LOSVD as described in
  Fig.~\ref{Bphase-space}. The two diamonds locate NGC~3311 and
  NGC~3309. The orange square shows the FoV used in the FORS2 MSIS
  observations. }
\label{gal-dist}
\end{figure*}

\begin{figure*}[hbt!] \centering
\includegraphics[width=6cm]{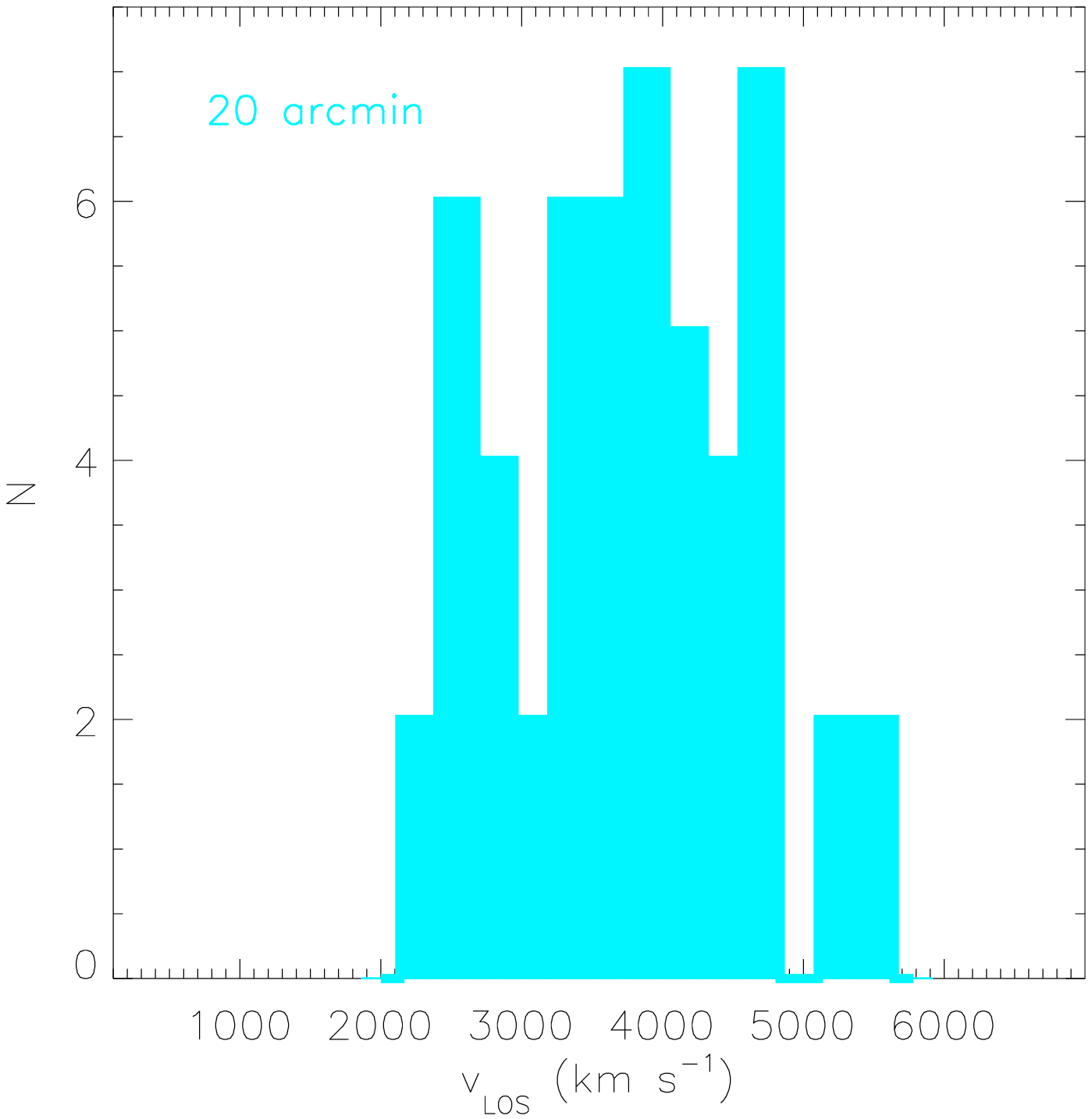}
\includegraphics[width=6cm]{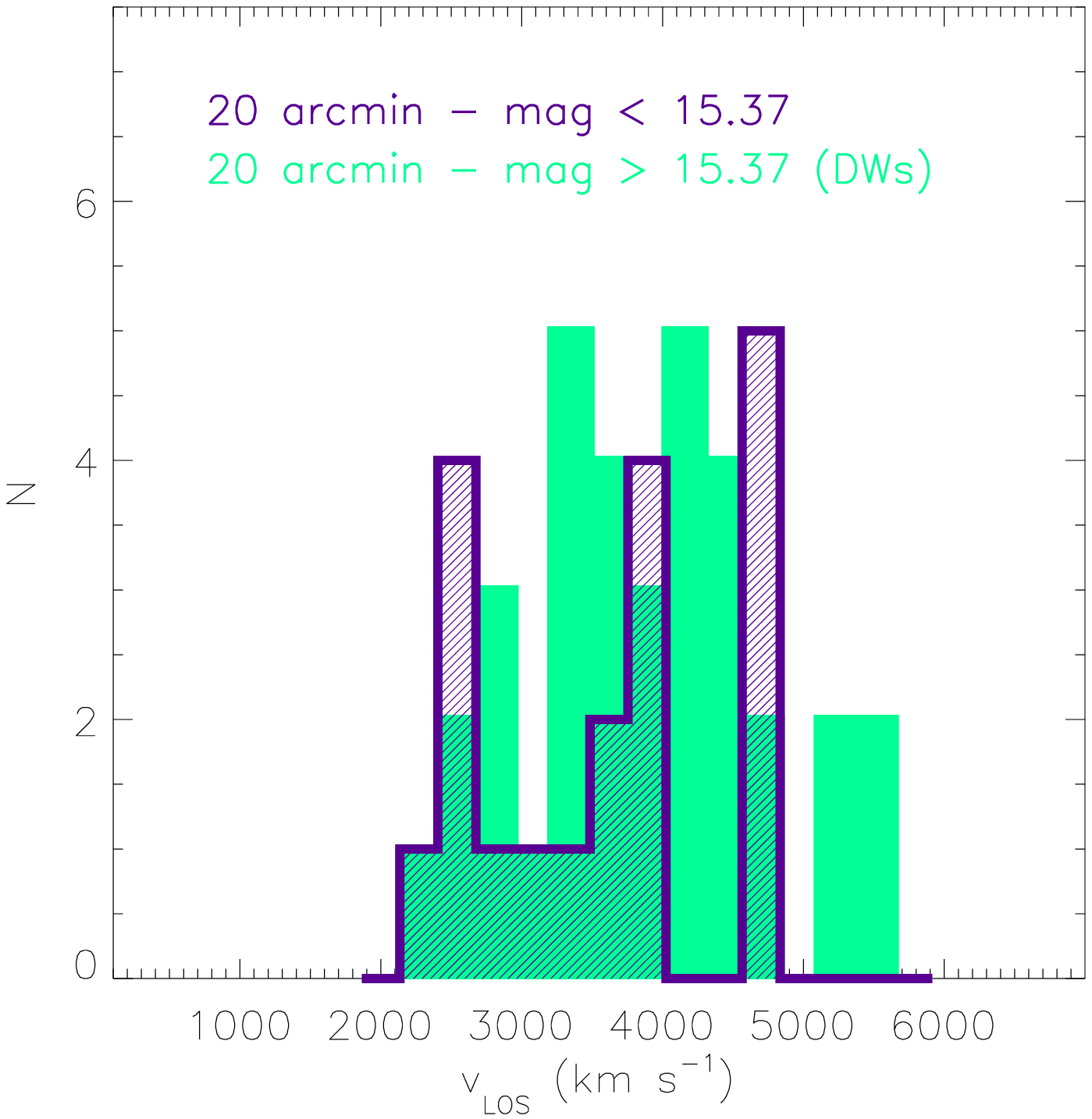}
\caption{\textit{Left panel}: Histogram showing the LOSVD of all
  galaxies from the catalog of \cite{Christlein03} within an area of
  $20\,\mbox{arcmin}$ in size centered around NGC~3311. \textit{Right
    panel}: The purple histogram indicates the LOSVD for the bright
  galaxies in this field, and the light green histogram the LOSVD for
  all dwarf galaxies from the catalog of \cite{Misgeld08} in the same
  area.}
\label{galaxies_histogram}
\end{figure*}

We now turn to a more general discussion of the spatial distribution
and kinematics of PNs and galaxies in the central region of the
cluster.  ICL is believed to originate from galaxies, so it is
interesting to ask whether the phase-space substructures seen in the
distribution of the PNs that trace the ICL has some correspondence to
similar structures in the distribution of cluster galaxies. Thus we
want to investigate the spatial distributions of the PNs associated
with the velocity subcomponents in the PN LOSVD discussed earlier, and
compare them with the spatial distribution of Hydra~I galaxies in
similar velocity intervals. In this way, we may obtain a better
understanding of the dynamical evolution of the galaxies in the
cluster core, and of the relevance of cluster substructures for the
origin of the diffuse cluster light.

\subsection{Spatial distributions of  the PN velocity components} \label{PNGalaxies1}

We first consider the spatial distribution of the PNs associated with
the different velocity components in the PN LOSVD.  This is shown in
the three panels of Fig.~\ref{Bphase-space}, divided according to the
classification in Sect.~\ref{modvsobs}.  Each panel covers a region of
$ 6.8 \times 6.8\, \mbox{arcmin}^2 \nsimeq
100\,\times100\,\mbox{kpc}^2$ centered on NGC~3311.

PNs associated with the central ICL component (middle panel of
Fig.~\ref{Bphase-space}) can be divided into two spatial
structures. There is a prominent PN group concentrated, as expected,
around NGC~3311, and an elongated east-west distribution in the
northern part of the FoV. By contrast, we see a low PN density region
in the southern part of the MSIS field.

Such a north/south asymmetry is seen also in the spatial distribution
of the galaxies.  Fig.~\ref{gal-dist} displays a larger area,
$20\,\times\,20\,\mbox{arcmin}^2$, which includes the MSIS field
studied in this work, as indicated by the orange square.  We can see
from the two panels (photo, and schematic) that NGC~3311 and NGC~3309
dominate the center of the MSIS field, that there is a high density of
bright galaxies in the northern part of the field, but a deficit of
galaxies to the south of NGC 3311.

The spatial distribution of the PNs associated with the secondary red
peak in the PN LOSVD is shown in the right panel of
Fig.~\ref{Bphase-space}.  It has a north/south elongation, apparently
extending further towards the south of NGC 3311 than the central ICL
component, with a high density region north/east of NGC~3311.

Finally, the spatial distribution of the PNs associated with the
secondary blue component at $1800\,\kms$ (left panel of
Fig.~\ref{Bphase-space}) also appears elongated along the
north/south direction, but the smaller number of objects in this
subsample makes inferring their spatial structure more difficult.

In summary, there is little evidence of a spherically symmetric
well-mixed distribution of PNs in the outer halo of NGC 3311 in
the cluster core. Several velocity components are seen, and even
the central ICL component centered on NGC 3311 shows signs of
spatial substructures.

\subsection{Spatial and velocity distribution of Hydra~I galaxies: 
comparison with the PNs sample} \label{Galaxies2}

The spatial distribution of the galaxies from
\citet{Christlein03,Misgeld08} in the central
$20\,\times\,20\,\mbox{arcmin}^2$ centered on NGC 3311 is shown in
Fig.~\ref{gal-dist}.  We would like to analyze their phase-space
distribution by dividing into the same velocity components as
identified in the PN LOSVD. Therefore, in the image on the left the
bright galaxies are encircled with the colors of the PN components in
Fig.~\ref{Bphase-space}, and in the right panel all galaxies in the
field are shown schematically as squares and crosses with the same
color code for these velocity bins. NGC~3311 and NGC~3309 are marked
in the center of the MSIS field (orange square).

In Fig.~\ref{galaxies_histogram}, the left panel shows the velocity
distribution of all the galaxies in the $20\,\times\,
20\,\mbox{arcmin}^2$ region centered on NGC~3311. In the right panel,
the velocity histograms for the bright galaxies ($\mbox{m}_R <
15.37$, violet color) and dwarf galaxies ($\mbox{m}_R > 15.37$,
green color) are shown separately.

The LOSVD for the Hydra~I galaxies covers the same velocities as for
the PN sample. If we select only galaxies in the range of velocities
of the PNs in the central ICL component, from $ 2800\,\kms$ to $
4450\,\kms$, their LOSVD is consistent with a Gaussian distribution
centered at a velocity of $3723\pm100\,\kms$ with a dispersion of
$542\pm80\,\kms$. This confirms results from long-slit kinematics in
the outer halo of NGC 3311 \citep{Ventimiglia10b}, where the velocity
dispersion was found to increase to $\nsim 465\,\kms$ at $\nsim 70"$
radius, $64\%$ of the velocity dispersion of all cluster galaxies.

This subsample of galaxies also has an interesting spatial
distribution: the central $6.8 \times 6.8\,\mbox{arcmin}^2$ region of
the cluster (the MSIS field), while dominated by NGC~3311 and
NGC~3309, contains no other Hydra~I galaxies with these
velocities. Whereas outside this region, they appear uniformly
distributed over the field (see the green squares and crosses in the
right panel of Fig.~\ref{gal-dist}). NGC~3311 is at the center of the
distribution of these galaxies both in space and in velocity.  The
distribution of these galaxies, as well as the similarity of their
velocity dispersion with that measured in the halo of NGC 3311,
supports the interpretation of \citet{Ventimiglia10b} that the halo of
NGC 3311 is dominated by intracluster stars that have been torn from
galaxies disrupted in the cluster core: galaxies that passed through
the central 100 kpc of the cluster core at modest velocities have all
been disrupted.

By contrast, the galaxies with LOS velocities $ > 4450\,\kms$ as in
the secondary red peak of the PN LOSVD are mostly located {\sl within}
the central $100 \times 100 $ kpc$^{2}$ region of the cluster (red
squares and crosses in the right panel of Fig.~\ref{gal-dist}). In
this subsample, there are 14 galaxies in total, 5 are classified as
bright galaxies and 9 are dwarfs, and 3 bright galaxies and 6 dwarfs
fall within the MSIS FORS2 field. These 6 dwarfs are concentrated in
the northeastern part of the halo of NGC~3311, in the same region
occupied by many PNs associated with the secondary red peak.

Finally, in this region there are only a few galaxies with a LOS
velocity lower than $2800\,\kms$, compatible with the secondary blue
peak in the PNs. They are 8 in total (blue squares and crosses in the
right panel of Fig.~\ref{gal-dist}).  Only one of these falls on the
boundary of the central $100 \times 100\, \mbox{kpc}^2$ region around
NGC~3311. One of these galaxies is the giant spiral NGC~3312,
south-east of NGC~3311. The others, including the spiral galaxy
NGC~3314, are located at larger distances from NGC 3311.

\subsection{Galaxy evolution and presence of substructures in the core
  of the Hydra~I cluster}

The distribution of galaxy properties in clusters holds important
information on galaxy evolution and the growth of galaxy clusters.  In
Sect.~\ref{Galaxies2}, we have discovered an apparent lack of galaxies
in the central $100 \times 100\, \mbox{kpc}^2$ region of the cluster
core with velocities in the same range as covered by the cD halo. A
similar result has been found in the NGC~5044 group
\citep{Mendel09}. A possible explanation is the tidal disruption of
galaxies at small cluster-centric radii. Galaxies with LOS velocities
in the range of the central ICL component of the Hydra I cluster are
no longer seen in the central region of the cluster, because they were
all disrupted in the past during close encounters with the luminous
galaxy and the dark matter distribution at the cluster center
\citep{Afalt05}. Their former stars now contributes to the diffuse
stellar component in the Hydra~I core.

Differently from the NGC~5044 group, however, we have found a number
of dwarf galaxies {\sl with high velocities} in the Hydra I core, with
small cluster-centric radii ($< 100$ kpc). These dwarfs have LOS
velocities higher than $4400\,\kms$ and seem to form a well defined
substructure both in velocity and spatial distribution. We speculate
that these galaxies are falling through the cluster core and are not
yet disrupted by the tidal interaction with NGC~3311. The PNLF of the
subsample of PNs associated spatially and in velocity with this
substructure places it almost at the same distance as the central ICL
component of the cluster: this group of galaxies may indeed now be on
the point of close encounter with NGC~3311 in the cluster center.

Finally, we have found a correlation between the PNs contributing to
the secondary blue peak of PN LOSVD, and 8 galaxies with a LOS
velocity lower than $2800\,\kms$. Among these galaxies is the large
spiral galaxy NGC~3312 ($\mbox{v}_{sys} = 2761\,\kms$) as well as
NGC~3314 ($\mbox{v}_{sys} = 2795\,\kms$). \cite{Fitchett88} and
\cite{McMahon92} have claimed the presence of a foreground group of
galaxies associated with these spirals.  Unfortunately due to the
small area covered by the current MSIS survey, it is difficult to
determine unambiguously whether the low velocity PNs (which from their
PNLF are at the distance of the cluster) are associated with these
galaxies. A PN survey covering the region between NGC~3311 and
NGC~3312 may provide a definite answer to this question.

\section{Summary and Conclusions} \label{Conclu}

Using multi-slit imaging spectroscopy with FORS2 on VLT-UT1, we have
studied a sample of 56 planetary nebula (PN) candidates in the Hydra~I
cluster at $50 {\rm Mpc}$ distance, targeting a region $100 \times
100\,\mbox{kpc}^2$ centered on the cluster cD galaxy, NGC~3311. The
MSIS technique allows us detect these emission sources and measure
their velocities, positions and magnitudes with a single observation.

PN candidates are defined as unresolved emission sources without
measurable continuum. Emission sources that are either resolved
spatially or in wavelength or have a detected continuum are classified
as background galaxies; see \citet{Ventimiglia10a}.  We show that the
luminosity function of the PN candidates is as expected for a
population of PNs at the distance of Hydra I. Moreover, almost all
the detected background galaxies occur in the velocity range between
$1000\,\kms$ and $2800\,\kms$, blue-shifted by $\gta 900\,\kms$
with respect to the mean recession velocity of the Hydra I cluster.
From these facts we conclude that any residual contamination of the
PN sample by background galaxies with undetectable continuum must
be small and restricted to the velocity range given.

The luminosity-specific number density $\alpha$ inferred from the PN
sample and the luminosity of diffuse light around NGC 3311 is
  a factor $\nsim 4$ lower than expected, even if we compare with
  the low $\alpha$ value determined from the (FUV-V) color which is
  one of the lowest for elliptical galaxies. A possible
interpretation is that ram pressure stripping by the dense, hot X-ray
emitting intracluster medium in the center of the cluster core around
NGC 3311 dramatically shortens the lifetime of the PN phase. This
also seems the most likely explanation for the observed lack of PNs
bound to NGC~3309, the other giant elliptical galaxy in the Hydra I
core.

The line-of-sight velocity distribution (LOSVD) of the observed PNs
shows at least three separate peaks, and their phase-space
distribution is inconsistent with a single well-mixed intracluster
distribution.  One peak, which we term the central intracluster
component, is broadly consistent with the outward continuation of the
intracluster halo of NGC 3311, which was earlier shown to have a
velocity dispersion of $\nsim470\,\kms$ at radii of $\gta50''$
\citep{Ventimiglia10b}. Simulating MSIS observations for a Gaussian
intrinsic LOSVD with $\nsim470\,\kms$ centered on the systemic velocity
of NGC~3311 has additionally shown significant residual asymmetries,
suggesting that also this central component is not completely
phase-mixed in the central cluster potential.

Many cluster galaxies are found in the LOS velocity range associated
with this central intracluster component ($2800\,\kms$ to $
4450\,\kms$), but {\sl none} in the central $100 \times
100\,\mbox{kpc}^2$ around NGC 3311. We suggest that the missing
galaxies have been disrupted by the gravitational field of NGC 3311
and the surrounding cluster dark matter, and that their light has been
added to the diffuse intracluster halo of NGC 3311 which is traced by
the PNs.

The second main peak in the PN LOSVD is centered at $5000\,\kms$, some
$1200\,\kms$ to the red of the main component.  In the same velocity
range, a number of dwarf galaxies are seen, which {\sl are} projected
onto the central $100 \times 100\,\mbox{kpc}^2$ around NGC 3311 where
also the PNs are located.  We suggest that the PNs and the galaxies in
this red peak of the LOSVD are linked, i.e., on similar orbits through
the cluster core, indicating that the galaxies have been partially
disrupted and the tidal debris is traced by the PNs. This will be the
subject of a further study based on deep photometry (Arnaboldi et al.\
2011, in preparation).

Finally, a third, blue peak in the PN LOSVD is seen at $\nsim
1800\,\kms$. The spatial distribution of these PNs is elongated in the
same sense as for the other two components in the cluster core, but
the number of sources with these velocities is smaller and a few of
them might be unresolved background galaxies. This makes it difficult
to establish a robust association between these PNs and cluster
galaxies, such as the group related to the spiral NGC~3312. A larger
survey area would be needed to establish such a link.

In summary, from this study of the kinematics of diffuse light in the
Hydra I cluster core with PNs, and the comparison with the projected
phase-space distribution of galaxies, we infer that: (1) The
intracluster stellar population in the Hydra~I cluster is not
well-mixed, even though this cluster is believed to be the prototype
of an evolved and dynamically relaxed cluster based on X-ray
indicators.  (2) The build-up of diffuse intracluster light and of the
cD halo of NGC 3311 are ongoing, through the accretion of material
from galaxies falling into the cluster core and tidally interacting
with its potential well.

\begin{acknowledgements}
  The authors thank the ESO VLT staff for their support during the
  MSIS observations. They also thank L.~Coccato, K.C.~Freeman and
  E.~Iodice for useful discussions.  This research has made use of the
  Gemini data archive and the NASA/IPAC Extragalactic Database (NED)
  operated by the Jet Propulsion Laboratory, California Institute of
  Technology.
\end{acknowledgements}

\bibliography{15982.bib}

\end{document}